\documentclass[10pt,journal,compsoc,letterpaper]{IEEEtran}

\ifCLASSOPTIONcompsoc
  \usepackage[nocompress]{cite}
\else
  \usepackage{cite}
\fi

\usepackage[pdftex]{graphicx}

\usepackage[cmex10]{amsmath}
\usepackage{amsfonts}
\usepackage{array}
\usepackage{fixltx2e}

\usepackage{color}
\usepackage{url}
\usepackage{booktabs}
\usepackage{comment}
\usepackage{multirow}

\usepackage[table]{xcolor}

\newcommand{\squishlist}{ 
   \begin{list}{$\bullet$}
    { \setlength{\itemsep}{0pt}      \setlength{\parsep}{3pt} 
      \setlength{\topsep}{3pt}       \setlength{\partopsep}{0pt}
      \setlength{\leftmargin}{1.5em} \setlength{\labelwidth}{1em}
      \setlength{\labelsep}{0.5em} } }

\newcommand{\squishend}{
  \end{list}  }

\usepackage{amsthm}
\theoremstyle{definition}
\newtheorem{definition}{Definition}
\newtheorem{example}{Example}
\newtheorem{theorem}{Theorem}

\usepackage{pifont}
\newcommand{\cmark}{\ding{51}}%
\newcommand{\xmark}{\ding{55}}

\usepackage{listings}

\usepackage{fancyhdr}
\begin{document}

\title{Efficient Parametric Model Checking Using Domain Knowledge}
\author{Radu~Calinescu, Colin~Paterson, and Kenneth~Johnson
\IEEEcompsocitemizethanks{\IEEEcompsocthanksitem R.~Calinescu and C.~Paterson are with the Department of Computer Science at the University of York, UK.
\IEEEcompsocthanksitem K.~Johnson is with the School of Engineering, Computer and Mathematical Sciences at the Auckland University of Technology, New Zealand.}}

\IEEEtitleabstractindextext{
\begin{abstract}
We introduce an efficient parametric model checking (ePMC) method for the analysis of reliability, performance and other quality-of-service (QoS) properties of software systems. ePMC speeds up the analysis of parametric Markov chains modelling the behaviour of software by exploiting domain-specific modelling patterns for the software components. To this end, ePMC precomputes closed-form expressions for key QoS properties of such patterns, and uses these expressions in the analysis of whole-system models. To evaluate ePMC, we show that its application to service-based systems and multi-tier software architectures reduces analysis time by several orders of magnitude compared to current parametric model checking methods. 
\end{abstract}

}

\maketitle

\thispagestyle{fancy}
\section{Introduction}

\emph{Parametric model checking} (PMC) \cite{Daws:2004:SPM:2102873.2102899,Hahn2011,Jansen2014} is a formal technique for the analysis of Markov chains with transitions probabilities specified as rational functions over a set of continuous variables. 
When the analysed Markov chains model software systems, these variables represent configurable parameters of the software or environment parameters unknown until runtime. The properties of Markov chains analysed by PMC are formally expressed in probabilistic computation tree logic (PCTL) \cite{Hansson1994} extended with rewards \cite{Andova2004}, and the results of the analysis are algebraic expressions over the same variables.

In software engineering, Markov chains are used to model the stochastic nature of software aspects including user inputs, execution paths and component failures, and the expressions generated by PMC correspond to reliability, performance and other quality-of-service (QoS) properties of the analysed software. The availability of algebraic expressions for these key QoS properties has multiple applications. First, evaluating the expressions for different parameter values enables the fast comparison of alternative system designs, e.g., in software product lines \cite{DBLP:conf/splc/GhezziS11,DBLP:journals/infsof/GhezziS13}. Second, self-adaptive software can efficiently evaluate the expressions at runtime, when the unknown environment parameters can be measured and suitable new values for the configuration parameters need to be selected  \cite{Filieri2011}. Third, PMC expressions allow the algebraic calculation of parameter values such that a QoS property satisfies a given constraint \cite{Daws:2004:SPM:2102873.2102899}. Finally, they enable the precise analysis of the sensitivity of QoS properties to changes in the system parameters  \cite{DBLP:journals/tse/FilieriTG16}.

PMC is supported by the model checkers PARAM \cite{Hahn2010}, PRISM \cite{prism} and Storm \cite{Dehnert2017}. However, despite significant advances in recent years \cite{Daws:2004:SPM:2102873.2102899,Hahn2011,Jansen2014}, the current PMC techniques (which these model checkers implement) are computationally very expensive, generate expressions that are often extremely large and inefficient to evaluate, and do not support the analysis of parametric Markov chains modelling important classes of software systems.  

Our work addresses these major limitations of existing PMC techniques and tools. To this end, we introduce an efficient parametric model checking (ePMC) method that exploits \emph{domain-specific modelling patterns}, i.e., ``fragments'' of parametric Markov chains occurring frequently in models of software systems from a domain of interest, and corresponding to typical ways of architecting software components within that domain.

\begin{figure}
\centering
\includegraphics[width=\hsize]{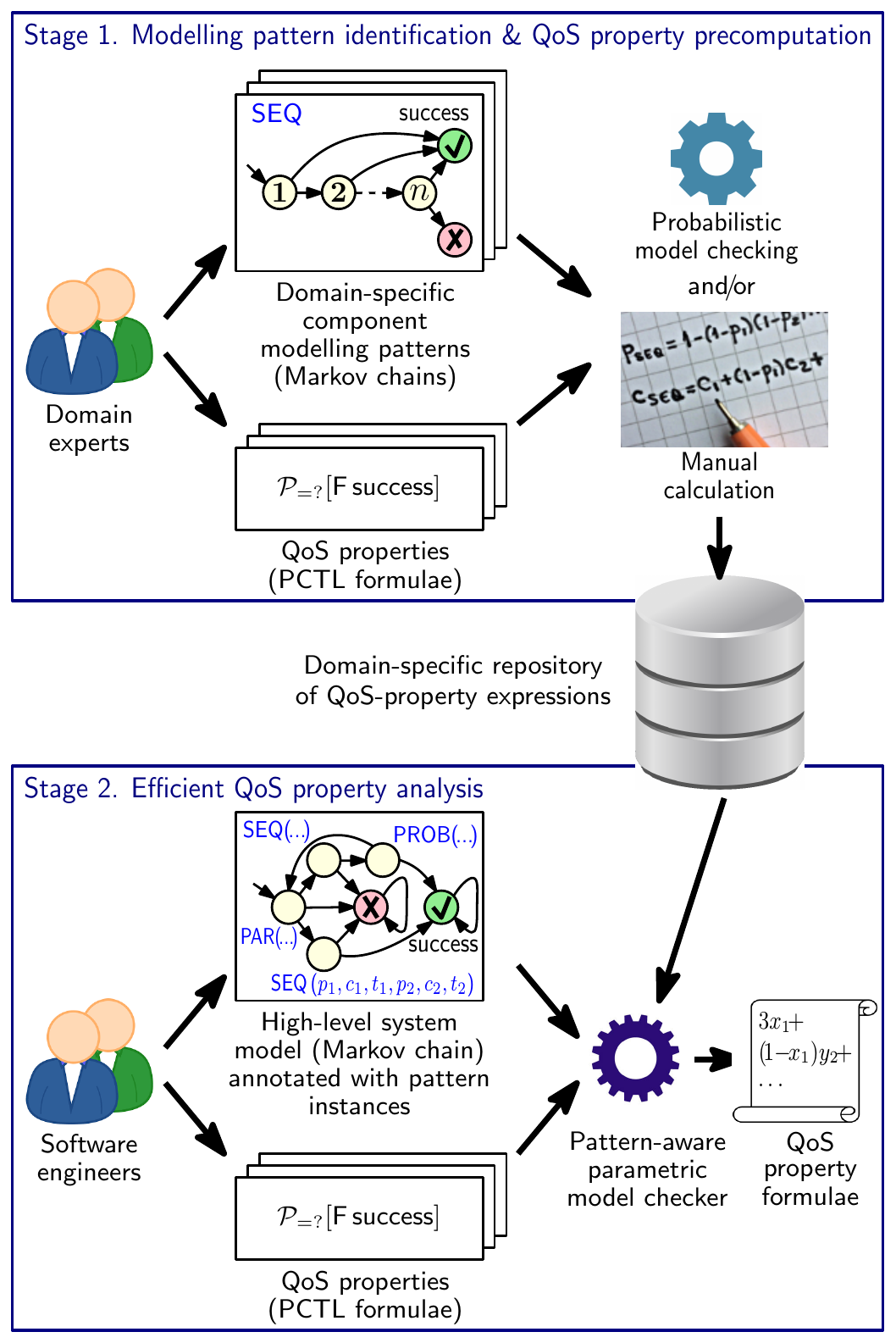}

\vspace*{-3mm}
\caption{Two-stage efficient parametric model checking \label{fig:method}}

\vspace*{-4mm}
\end{figure}

As shown in Fig.~\ref{fig:method}, ePMC comprises two stages. The first stage is 
performed only once for each domain that ePMC is applied to. This stage uses domain-expert input to identify modelling patterns for  components of systems from the considered domain, and precomputes closed-form expressions for key QoS properties of these patterns. 
For example, the modelling patterns for the service-based systems domain (described in detail in Section~\ref{sect:sbs}) correspond to different ways in which $n\geq 1$ functionally-equivalent services can be used to execute an operation of the system. One option is to invoke the $n$ services \emph{sequentially}, such that service~1 is always invoked, and service $i\!>\!1$ is only invoked if the invocations of services 1, 2, \ldots, $i-1$ have all failed. The component modelling pattern labelled `$\mathsf{SEQ}$' at the top of Fig.~\ref{fig:method} depicts this option. The graphical representation of the pattern shows the invocations of the $n$ services as states labelled $1$, $2$, \ldots, $n$, and the successful and failed completion of the operation as states labelled with a tick '\cmark' and a cross '\xmark', respectively. QoS properties such as the probability of reaching the success state and the expected execution time and \mbox{cost of the operation for this pattern can be computed as}\\[1.5mm] 
\hspace*{5mm}$\mathit{prob}=p_1+(1-p_1)p_2+\ldots+\bigl(\prod_{i=1}^{n-1}(1-p_i)\bigr)p_n$\\[1.5mm]
\hspace*{5mm}$\mathit{time}=t_1+(1-p_1)t_2+\ldots+\bigl(\prod_{i=1}^{n-1}(1-p_i)\bigr)t_n$\\[1.5mm]
\hspace*{5mm}$\mathit{cost}=c_1+(1-p_1)c_2+\ldots+\bigl(\prod_{i=1}^{n-1}(1-p_i)\bigr)c_n$\\[1.5mm]
where $p_i$, $t_i$ and $c_i$ are the probability of successful invocation, the execution time and the cost of service $i$, $1\leq i\leq n$, respectively. 
As illustrated in Fig.~\ref{fig:method}, these calculations can be carried out using an existing probabilistic model checker  
or manually. The resulting expressions are stored in a domain-specific repository, and are used in the next ePMC stage. 

The second ePMC stage is performed for each structurally different variant of a system and QoS property under analysis. The stage involves the PMC of a parametric Markov chain that models the interactions between the system components. This Markov model can be provided by software engineers with PMC expertise, or can be generated from more general software models, such as UML activity diagrams annotated with probabilities as in \cite{6693145,Gallotti:2008:QPS:1478067.1478078,7968147}. The model states associated with system components are labelled with \emph{pattern instances} that specify the modelling pattern used for each component and its parameters. For instance, the pattern instance $\mathsf{SEQ}(p_1,c_1,t_1,p_2,c_2,t_2)$ from Fig.~\ref{fig:method} labels a component implemented using the sequential pattern described earlier and $n=2$ services with success probabilities $p_1,p_2$, costs $c_1,c_2$ and mean execution times $t_1,t_2$. The pattern-annotated Markov model is analysed by a model checker with pattern manipulation capabilities. The result of the analysis is a set of formulae comprising:
\squishlist
\item A formula for the system-level QoS property, specified as a function over the component-level QoS property values. This formula is obtained by applying standard PMC to the pattern-annotated Markov model;
\item Formulae for the relevant component-level QoS properties. These formulae are obtained by instantiating the appropriate closed-form expressions from the domain-specific repository produced in the first ePMC stage.
\squishend
All ePMC formulae are rational functions that can be efficiently evaluated for any combinations of parameter values, e.g., using tools such as Matlab and Octave. 

The main contributions of our paper are: 
\squishlist
\item[1)] A theoretical foundation for the ePMC method.
\item[2)] An open-source tool that automates the application of the method, and is freely available from our project website \url{https://www.cs.york.ac.uk/tasp/ePMC/}.
\item[3)] Repositories of modelling patterns for the service-based systems and multi-tier software architecture domains.
\item[4)] An extensive evaluation which shows that ePMC is several orders of magnitude faster and produces much smaller algebraic expressions compared to the PMC techniques currently implemented by the leading model checkers PARAM, PRISM and Storm, in addition to supporting the analysis of parametric Markov chains that are too large for these model checkers. 
\squishend
These contributions build on our preliminary work from \cite{DBLP:conf/icse/CalinescuJP18}, extending it with a theoretical foundation, tool support, repositories of modelling patterns for two domains, and a significantly larger evaluation.

The rest of the paper is structured as follows. Section~\ref{sect:preliminaries} provides a brief introduction to the model checking of parametric Markov chains. Section~\ref{sect:running} describes a simple service-based system that we then use as a running example when presenting the ePMC theoretical foundation in Section~\ref{sect:theory}. Section~\ref{sect:tool} covers the implementation of the ePMC tool, while Sections~\ref{sect:sbs} and~\ref{sect:multitier} detail the application of ePMC to the service-based systems and multi-tier software architectures domains, respectively. Section~\ref{sect:evaluation} presents our experimental results, and Section~\ref{sect:related} compares our method with related work. Finally, Section~\ref{sect:conclusion} provides a brief summary and discusses our plans for future work.

\section{Preliminaries \label{sect:preliminaries}}

\subsection{Parametric Markov chains \label{sect:prelim-1}}

\emph{Markov chains} (MCs) are finite state transition systems used to model the stochastic behaviour of real-world systems. MC states correspond to relevant configurations of the modelled system, and are labelled with atomic propositions which hold in those states. State transitions model all possible transitions between states, and are annotated with probabilities as specified by the following definition.

\begin{definition}
A Markov chain $M$ over a set of atomic propositions $\mathit{AP}$ is a tuple 
\begin{equation}
  M=(S,s_0,\mathbf{P},L),
  \label{eq:DTMC}
\end{equation} 
where $S$ is the finite set of MC states; $s_0\in S$ is the initial state; $\mathbf{P}: S\times S\to [0,1]$ is a transition probability matrix where, for any states $s,s'\in S$, $\mathbf{P}(s,s')$ is the probability of transitioning to state $s$ from state $s'$; and $L : S \to 2^{AP}$ is the state labelling function. 
\end{definition}

A state $s$ of a Markov chain $M$ is an \emph{absorbing state} if $\mathbf{P}(s, s) \!=\! 1$ and $\mathbf{P}(s, s') \!=\! 0$ for all $s'\!\neq\! s$, and a \emph{transient state} otherwise. A \emph{path} $\pi$ over $M$ is a possibly infinite sequence of states from $S$ such that for any adjacent states $s$ and $s'$ in $\pi$, $\mathbf{P}(s,s')>0$. The $m$-th state on a path $\pi$, $m\geq 1$, is denoted $\pi(m)$. For any state $s$, $\mathit{Paths}^M(s)$ represents the set of all infinite paths over $M$ that start with state $s$. Finally, we assume that every state $s\in S$ is reachable from the initial state, i.e., there exists a path $\pi\!\in\! \mathit{Paths}^M(s_0)$ such that $\pi(i)=s$ for some $i>0$.

To compute the probability that a Markov chain~(\ref{eq:DTMC}) behaves in a specified way when in state $s$, we use a \emph{probability measure} $\mathrm{Pr}_s$ defined over $\mathit{Paths}^M(s)$ such that \cite{kemeny-etal1976,BaierKatoen2008}:
\begin{equation*}
  \begin{array}{l}
  \!\!\mathrm{Pr}_s(\{\pi\!\in\! \mathit{Paths}^M\!(s)\!\mid\! \pi\!=\!s_{1}s_{2}\ldots s_{m}\ldots\}) \!=\\ \qquad\qquad\qquad\mathbf{P}(s_{1}s_{2}\ldots s_{m})=\prod_{i=1}^{m-1}\! \mathbf{P}(s_i,s_{i+1}), 
  \end{array}
\end{equation*}
\mbox{where $\{\pi\!\in\! \mathit{Paths}^M\mid \pi=s_{1}s_{2}\ldots s_{m}\ldots\}$ is the set of all} infinite paths that start with the prefix $s_{1}s_{2}\ldots s_{m}$ (i.e., the \emph{cylinder set} of this prefix). Further details about this probability measure and its properties are available from \cite{kemeny-etal1976,BaierKatoen2008}.

To allow the verification of a broader set of QoS properties, MC states  
can be annotated with nonnegative values termed \emph{rewards} \cite{Andova2004}. These values are interpreted as ``costs'' (e.g.\ energy used) or ''gains'' (e.g.\ requests processed). 

\begin{definition}
A \emph{reward structure} over a Markov chain $M=(S,s_0,\mathbf{P},L)$ is a function $\rho:S\to\mathbb{R}_{\geq 0}$. For any state $s\in S$, $\rho(s)$ represents the reward ``earned'' on leaving state $s$.
\end{definition}

Our work focuses on the analysis of parametric Markov chains (sometimes called \emph{incomplete Markov chains} \cite{BenediktLW2013} or \emph{uncertain Markov chains} \cite{SenVA2006}).

\begin{definition}
A \emph{parametric Markov chain} is an MC comprising transition probabilities $\mathbf{P}(s,s')$ and/or rewards $\rho(s)$ defined as rational functions over a set of continuous variables \cite{Daws:2004:SPM:2102873.2102899,Hahn2011,Jansen2014}.  
\end{definition}

\subsection{Property specification \label{sect:prelim-2}}

The properties of Markov chains are formally expressed in probabilistic variants of temporal logic. In our work we use probabilistic computation tree logic (PCTL) \cite{Ciesinski2004,Hansson1994} extended with rewards \cite{Andova2004}, which is supported by all leading probabilistic model checkers.  Rewards-extended PCTL allows the specification of probabilistic and reward properties using the probabilistic operator $\mathcal{P}_{\bowtie p}[\cdot]$ and the reward operator $\mathcal{R}_{\bowtie r}[\cdot]$, respectively, where $p\in [0,1]$ is a probability bound, $r\in \mathbb{R}_{\geq 0}$ is a reward bound, and $\bowtie\,\in $ $\{ \geq, >, <, \leq\}$ is a relational operator. Formally, a \emph{state formula} $\Phi$ and a \emph{path formula} $\Psi$ in PCTL are defined by the grammar:
\begin{align}
& \Phi::=  true \;\vert\; a \;\vert\; \Phi \wedge \Phi \;\vert\; \neg \Phi \;\vert\; \mathcal{P}_{\bowtie p} [\Psi]
\label{eq:pctl} \\
& \Psi::= X \Phi \;\vert\; \Phi\; \mathrm{U}\; \Phi \;\vert\; \Phi\; \mathrm{U}^{\leq k}\, \Phi
\label{eq:pctl2}
\end{align}
and a \emph{reward state formula} is defined by the grammar:
\begin{equation}
  \Phi::=  \mathcal{R}_{\bowtie r} [\mathrm{I}^{=k}] \;\vert\;
               \mathcal{R}_{\bowtie r} [\mathrm{C}^{\leq k}] \;\vert\;
               \mathcal{R}_{\bowtie r} [\mathrm{F}\; \Phi] \;\vert\;
               \mathcal{R}_{\bowtie r} [\mathrm{S}],
\label{eq:pctl3}
\end{equation}
where $k\!\in\! \mathbb{N}_{>0}$ is a timestep bound and $a\!\in\! AP$ is an atomic proposition. 

The PCTL semantics is defined using a satisfaction relation $\models$ over the states $S$ and the paths $\mathit{Paths}^M(s)$, $s\in S$, of a Markov chain~(\ref{eq:DTMC}). Given a state $s$ and a path $\pi$ of the Markov chain, $s\models \Phi$ means ``$\Phi$ holds in state $s$'',  $\pi\models \Psi$ means ``$\Psi$ holds for path $\pi$'', and we have: 
\squishlist
\item[--] $s\models true$ for all $s\in S$; 
\item[--] $s \models a$ iff $a\in L(s)$; 
\item[--] $s \models \neg \Phi$ iff $\neg (s\models \Phi)$; 
\item[--] $s\models \Phi_1 \wedge \Phi_2$ iff $s\models \Phi_1$ and $s\models \Phi_2$;
\item[--] $s\models \mathcal{P}_{\bowtie p} [\Psi]$ iff $\mathrm{Pr}_s(\{\pi\in \mathit{Paths}^M(s) \mid \pi \models \Psi\}) \bowtie p$.
\item[--] the \emph{next path formula} $X \Phi$ holds for path $\pi$  
iff $\pi(2)\models \Phi$;
\item[--] the \emph{time-bounded until path formula} $\Phi_1\, \mathrm{U}^{\leq k}\, \Phi_2$ holds for path $\pi$  
iff $\Phi_2$ holds in the $i$-th path state, $i\leq k$, and $\Phi_1$ holds in the first $i-1$ path states, i.e.:
\[
   \exists i\leq k.(\pi(i)\models \Phi_2 \wedge \forall j<i.\pi(j)\models \Phi_1).
\]
\item[--] the \emph{unbounded until formula} $\Phi_1\,\mathrm{U}\, \Phi_2$ removes the bound $k$ from the  time-bounded ``until'' formula.
\squishend
The notation $\mathrm{F}\,\Phi\equiv \mathit{true} \,\mathrm{U}\, \Phi$ is used when the first part of an until formula is $\mathit{true}$. Thus, the \emph{reachability property} $\mathcal{P}_{\bowtie p} [\mathrm{F}\,\Phi]$ holds if the probability of reaching a state where $\Phi$ is true satisfies $\bowtie p$.
Finally, the reward state formulae specify the expected values for: the \emph{instantaneous reward} at timestep $k$, $\mathcal{R}_{\bowtie r} [\mathrm{I}^{=k}]$; the \emph{cumulative reward} up to timestep $k$, $\mathcal{R}_{\bowtie r} [\mathrm{C}^{\leq k}]$; the \emph{reachability reward} cumulated until reaching a state that satisfies a property $\Phi$, $\mathcal{R}_{\bowtie r} [\mathrm{F}\,\Phi]$; and the \emph{steady-state reward} in the long run, $\mathcal{R}_{\bowtie r} [\mathrm{S}]$. For a detailed description of the PCTL semantics, see \cite{Ciesinski2004,Hansson1994,Andova2004}.

\subsection{Parametric model checking}

Probabilistic model checkers including  MRMC \cite{Katoen:2011:IOP:1930080.1930191}, PRISM  \cite{prism} and Storm \cite{Dehnert2017} support the verification of PCTL properties of Markov chains. To verify whether a formula $\mathcal{P}_{\bowtie p} [\Psi]$ holds in a state $s$, these tools first compute the probability $p'$ that $\Psi$ holds for MC paths starting at $s$, and then compare $p'$ to the bound $p$. The actual probability $p'$ can also be returned (for the outermost operator $\mathcal{P}$ of a formula) so PCTL was extended to include the formula $\mathcal{P}_{=?} [\Psi]$ denoting this probability. Likewise, the extended-PCTL formulae $\mathcal{R}_{=?} [\mathrm{I}^{=k}]$, $\mathcal{R}_{=?} [\mathrm{C}^{\leq k}]$, $\mathcal{R}_{=?} [\mathrm{F}\;\Phi]$, and $\mathcal{R}_{=?} [\mathrm{S}]$ denote the actual values of the expected rewards from~(\ref{eq:pctl3}).

\emph{Parametric model checking} (PMC) represents the verification of \emph{quantitative PCTL properties} $\mathcal{P}_{=?} [.]$ without nested probabilistic operators and  reward properties $\mathcal{R}_{=?} [.]$ of parametric Markov chains using algorithms such as \cite{Daws:2004:SPM:2102873.2102899,Hahn2011,Jansen2014}. The PMC verification result is a rational function of the variables used to define the transition probabilities of the verified parametric Markov chain. PMC is supported by verification tools including the dedicated model checker PARAM \cite{Hahn2010}, the latest versions of PRISM \cite{prism}, and the recently released model checker Storm \cite{Dehnert2017}.

\section{Running Example \label{sect:running}}

\begin{figure}
\centering
\includegraphics[width=0.96\hsize]{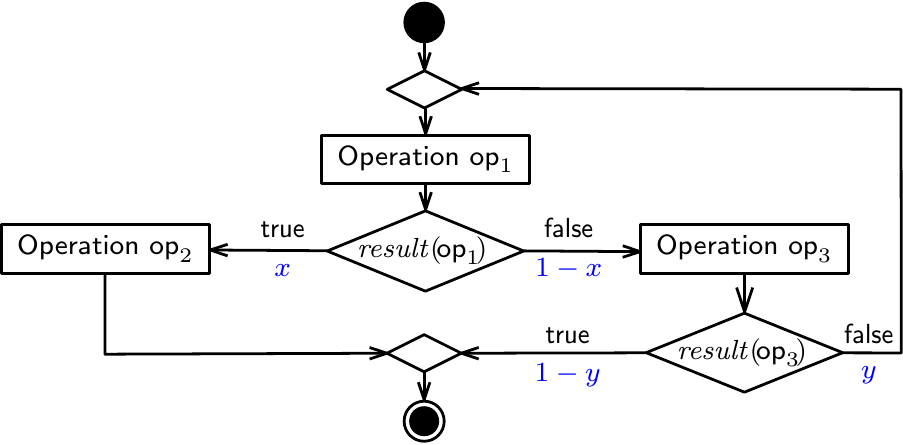}

\vspace*{-2mm}
\caption{UML activity diagram of the system used as a running example \label{fig:running}}
\end{figure}

We will illustrate the theoretical aspects and the application of our ePMC method using a service-based system that implements the simple workflow from Fig.~\ref{fig:running}. This workflow handles user requests by first performing an operation $\mathsf{op}_1$.  Depending on the result of $\mathsf{op}_1$, its execution is followed by the execution of either operation $\mathsf{op}_2$ or operation $\mathsf{op}_3$. The execution of $\mathsf{op}_2$ completes the workflow, while after the execution of $\mathsf{op}_3$ the workflow may terminate or may need to re-execute $\mathsf{op}_1$. The outgoing branches of the decision nodes from Fig.~\ref{fig:running} are annotated with their unknown probabilities of execution ($x$ and $1-x$, and $y$ and $1-y$).

\begin{figure*}
\centering
\includegraphics[width=0.95\hsize]{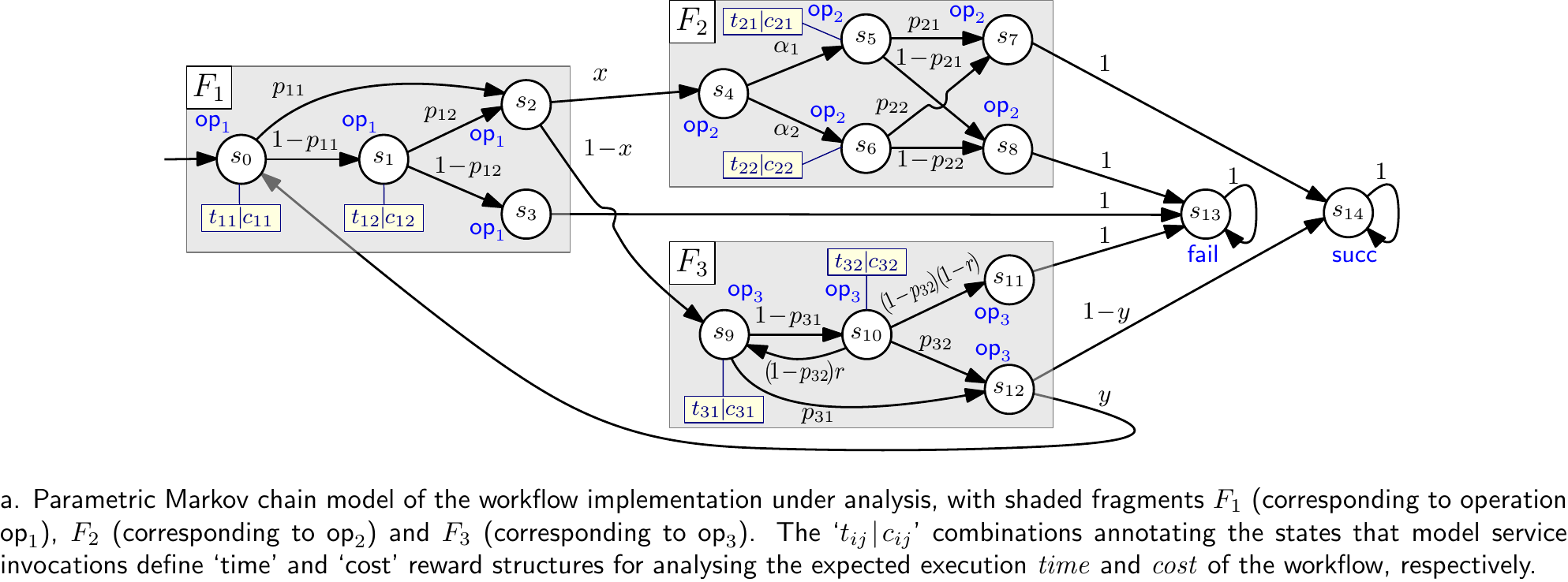}

\vspace*{5mm}
\includegraphics[width=0.95\hsize]{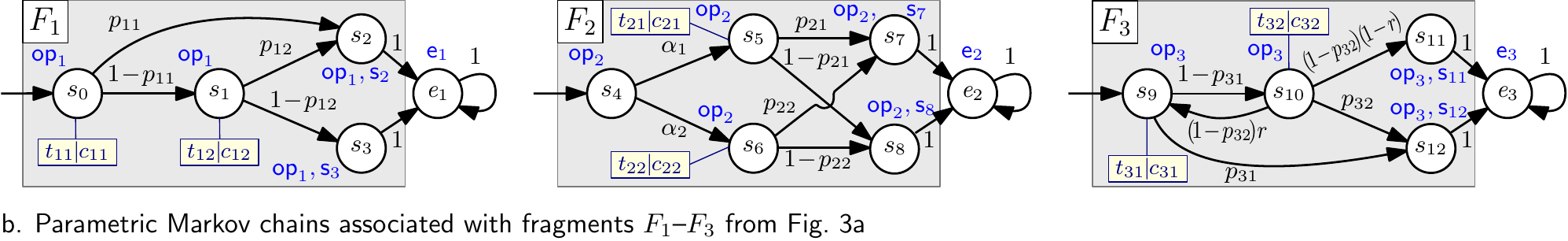}

\vspace*{5mm}
\includegraphics[width=0.95\hsize]{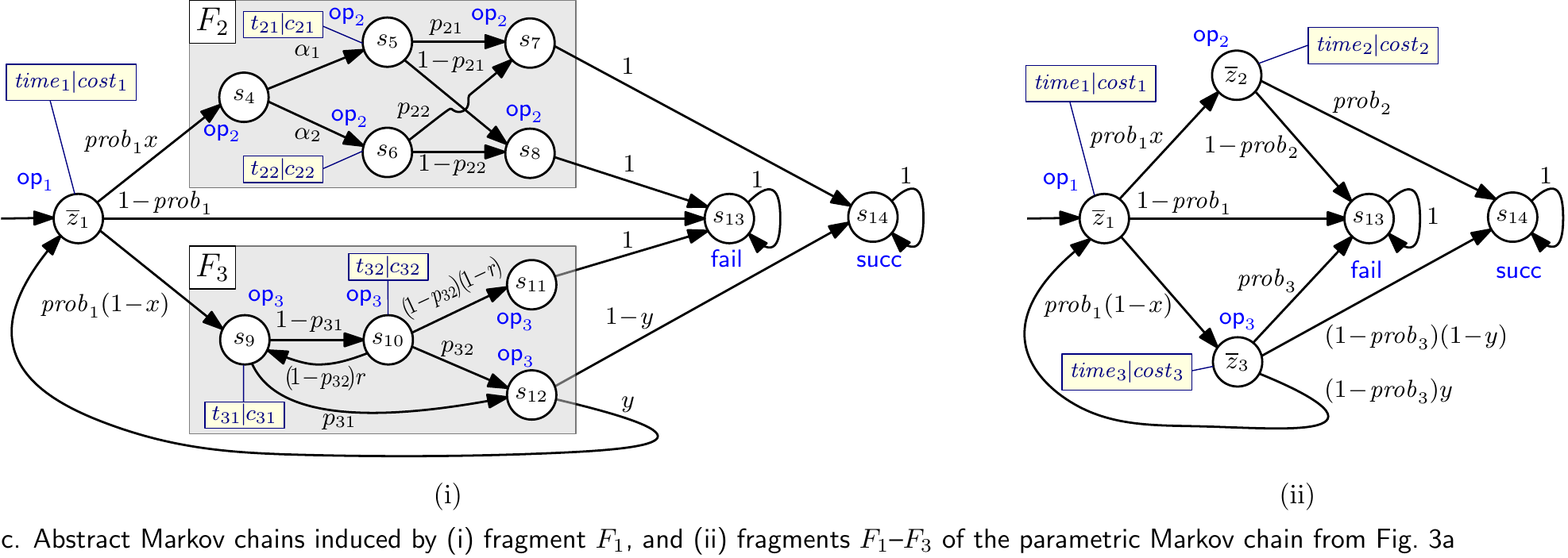}
\caption{ePMC application to a parametric Markov chain model of the workflow from the running example \label{fig:running-pmc}}
\end{figure*}

We suppose that multiple functionally-equivalent services $\mathsf{svc}_{i1}$, $\mathsf{svc}_{i2}$, \ldots can be used to perform each  operation $\mathsf{op}_i$, $i\in\{1,2,3\}$, and that these services have probabilities of successful invocation $p_{i1}, p_{i2}, \ldots$, expected response times $t_{i1}, t_{i2}, \ldots$ and invocation costs $c_{i1}, c_{i2}, \ldots$ Accordingly, the workflow can be implemented using different system architectures and service combinations. 
Our running example considers the implementation where:
\squishlist
\item Operation $\mathsf{op}_1$ is executed by invoking services $\mathsf{svc}_{11}$ and $\mathsf{svc}_{12}$ \emph{sequentially}, such that service $\mathsf{svc}_{11}$ is always invoked, and service $\mathsf{svc}_{12}$ is only invoked if the invocation of $\mathsf{svc}_{11}$ fails (i.e., times out or returns an error). As a result, the operation completes successfully whenever either service invocation is successful, and fails when the invocations of both services fail.
\item Operation $\mathsf{op}_2$ is executed using services $\mathsf{svc}_{21}$ and $\mathsf{svc}_{22}$ \emph{probabilistically}, such that $\mathsf{svc}_{2j}$ is invoked with probability $\alpha_{j}$ for $j\in\{1,2\}$, where $\alpha_{1}+\alpha_{2}=1$.  
\item Operation $\mathsf{op}_3$ is executed by invoking services $\mathsf{svc}_{31}$ and $\mathsf{svc}_{32}$ \emph{sequentially with retry}. This involves invoking the two services sequentially (as for $\mathsf{op}_1$) and, if both service invocations fail, retrying the execution of the operation by using the same strategy with probability $r\in(0,1)$. 
\squishend
The parametric Markov chain from Fig.~\ref{fig:running-pmc}a models this implementation of the workflow. For instance, the MC states $s_0$ and $s_1$ (labelled `$\mathsf{op}_1$') model the execution of operation $\mathsf{op}_1$ by first invoking service $\mathsf{svc}_{11}$ (state $s_0$) and, if service service $\mathsf{svc}_{11}$ fails (which happens with probability $1-p_{11}$), also invoking $\mathsf{svc}_{12}$ (state $s_1$). The invocation of $\mathsf{svc}_{12}$ fails with probability $1-p_{12}$, in which case the system transitions to state $s_3$ and then to the `\textsf{fail}' state $s_{13}$. If either $\mathsf{svc}_{11}$ or $\mathsf{svc}_{12}$ succeeds (state $s_2$), there is a probability $x$ that operation $\mathsf{op}_2$ (modelled by states $s_4$--$s_8$) is executed next, and a probability $1-x$ that the next operation is $\mathsf{op}_3$ (modelled by states $s_9$--$s_{12}$). 

To model the execution of operation $\mathsf{op}_2$, the MC includes transitions with probabilities $\alpha_1$ and $\alpha_2$ from $s_4$ to state $s_5$ (which corresponds to the invocation of service $\mathsf{svc}_{21}$) and to state $s_6$ (which corresponds to the invocation of service $\mathsf{svc}_{22}$), respectively. The successful execution of $\mathsf{svc}_{21}$ or $\mathsf{svc}_{22}$ results in state $s_7$ being reached and in the successful completion of the workflow (state $s_{14}$, labelled `\textsf{succ}'), while a failed invocation results in state $s_8$ being reached and in the failure of the workflow (state $s_{13}$). 

Finally, states $s_{9}$--$s_{12}$ model the execution of operation $\mathsf{op}_3$ similarly to how $\mathsf{op}_1$ is modelled by $s_0$--$s_3$, except that a successful execution of $\mathsf{op}_3$ is followed by $\mathsf{op}_1$ (with probability $y$) or the successful end of the workflow (with probability $1-y$), and failed invocations of $\mathsf{svc}_{32}$ lead to a retry of the operation (with probability $r$) or to the failure of $\mathsf{op}_3$ and thus of the entire workflow (with probability $1-r$).

Parametric model checking applied to the MC from Fig.~\ref{fig:running-pmc}a can compute closed-form expressions for a wide range of QoS properties of the system. These properties can then be evaluated very efficiently for different combinations of services with different parameters. For our running example, we assume that the software engineers developing the system are interested to analyse the following properties:
\begin{enumerate}
\item[(i)] The probability $P_1$ that the workflow implemented by the system completes successfully;
\item[(ii)] The probability $P_2$ that the workflow fails due to a failed execution of operations $\mathsf{op}_1$ or $\mathsf{op}_2$;
\item[(iii)] The expected execution time $T$ of the workflow;
\item[(iv)] The expected cost $C$ of executing the workflow.
\end{enumerate}
Table~\ref{tab:running-example-properties} presents these properties formalised in PCTL, and their closed-form expressions computed using the probabilistic model checker Storm and significantly simplified through manual factorisation. Although PMC is feasible for the simple model from Fig.~\ref{fig:running-pmc}, the complex expressions from Table~\ref{tab:running-example-properties} already suggest that this might not be true for larger systems and models. The experimental results presented later in Section~\ref{sect:evaluation} confirm that indeed the PMC techniques implemented by current model checkers do not scale to much larger systems than the one from Fig.~\ref{fig:running}---a limitation addressed by our ePMC method described next. 

\begin{table*}
\caption{Parametric model checking of the four QoS properties from the running example}
\label{tab:running-example-properties}
\centering
\renewcommand{\arraystretch}{1.5}
\begin{tabular}{p{0.4cm}p{2.1cm}p{14.5cm}} 
\toprule
\textbf{Prop.} & \textbf{PCTL formula} & \textbf{PMC expression} \\
\midrule 
$P_1$ & $\mathcal{P}_{=?}[\mathsf{F\; succ}]$ & 
$
\bigl\{
p_{11}p_{12}\bigl[
x(\alpha_2p_{31}p_{22}p_{32}r
-\alpha_1p_{31}p_{21}r
+\alpha_1p_{21}r
-\alpha_1p_{21}p_{32}r
-\alpha_2p_{22}
+\alpha_1p_{31}p_{21}p_{32}r
-yp_{32}
+p_{31}
-p_{31}p_{32}
+p_{31}yp_{32}
-\alpha_2p_{22}p_{32}r
+\alpha_2p_{22}r
+p_{32}
-p_{31}y
-\alpha_2p_{31}p_{22}r
-\alpha_1p_{21}
)
\bigr]
+p_{11}\bigl[
x(
\alpha_1(p_{21}-p_{31}p_{21}p_{32}r+p_{21}p_{32}r-p_{21}r+p_{31}p_{21}r)
+\alpha_2(p_{31}p_{22}r-p_{31}p_{22}p_{32}r-p_{22}r+p_{22}p_{32}r+p_{22})
)
\bigr]
+p_{12}\bigl[
x(
\alpha_1(p_{21}p_{32}r-p_{21}r-p_{31}p_{21}p_{32}r+p_{21}+p_{31}p_{21}r)
+\alpha_2(p_{22}+p_{22}p_{32}r-p_{22}r+p_{31}p_{22}r-p_{31}p_{22}p_{32}r)
)
+(p_{11}p_{12}-(p_{11}+p_{12})(1-x)) (p_{31}p_{32}-p_{31}-p_{32})(1-y)
\bigr]
\bigr\}
/
\bigl\{
\bigl[
(1-x)y(p_{11}+p_{12}-p_{12}p_{12})-r
\bigr]
(p_{31}p_{32}-p_{31}-p_{32})
+1
-r
\bigr\}
$
\\
$P_2$ & $\mathcal{P}_{=?}[\neg \mathsf{op}_3\; \mathrm{U}\; \mathsf{fail}]$ & 
$
p_{11}\bigl[
x(\alpha_1(p_{12}p_{21}-p_{12}+1-p_{21})+\alpha_2(p_{12}p_{22}-p_{12}+1-p_{22}))
+p_{12}-1
\bigr]
+p_{12}\bigl[
x(\alpha_1(1-p_{21})+\alpha_2(1-p_{22}))-1
\bigr]
+1
$
\\
$\mathit{T}$ & $\mathcal{R}^\mathsf{time}_{=?}[\mathsf{F\; succ \vee fail}]$ & 
$
\bigl\{
p_{11}p_{12}\bigl[
x(
\alpha_1(t_{21}r-t_{21}-t_{21}p_{32}r-p_{31}t_{21}r+p_{31}t_{21}p_{32}r)
+\alpha_2(p_{31}t_{22}p_{32}r-t_{22}+t_{22}r-t_{22}p_{32}r-p_{31}t_{22}r)
+t_{32}
+t_{31}
-p_{31}t_{32}
)
+t_{12}p_{31}p_{32}r
-t_{32}
-t_{31}
+p_{31}t_{32}
-t_{12}
+t_{12}r
-t_{12}p_{32}r
-t_{12}p_{31}r
\bigr]
+p_{11}\bigl[
x(
\alpha_2(t_{22}-t_{22}r+t_{22}p_{32}r+p_{31}t_{22}r-p_{31}t_{22}p_{32}r)
+\alpha_1(t_{21}-t_{21}r+t_{21}p_{32}r+p_{31}t_{21}r-p_{31}t_{21}p_{32}r)
-t_{32}
-t_{31}
+p_{31}t_{32}
)
+t_{32}
+t_{31}
-p_{31}t_{32}
-t_{12}r
+t_{12}p_{32}r
+t_{12}p_{31}r
-t_{12}p_{31}p_{32}r
-t_{11}
+t_{11}r
-t_{11}p_{32}r
-t_{11}p_{31}r
+t_{11}p_{31}p_{32}r
+t_{12}
\bigr]
+p_{12}\bigl[
x(
\alpha_1(t_{21}p_{32}r+p_{31}t_{21}r-p_{31}t_{21}p_{32}r+t_{21}-t_{21}r)
+\alpha_2(t_{22}-t_{22}r+t_{22}p_{32}r+p_{31}t_{22}r-p_{31}t_{22}p_{32}r)
-t_{32}
-t_{31}
+p_{31}t_{32}
)
+t_{31}
-p_{31}t_{32}
+t_{32}
+t_{12}
-t_{12}r
+t_{12}p_{32}r
+t_{12}p_{31}r
-t_{12}p_{31}p_{32}r
\bigr]
+t_{11}
-t_{11}r
+t_{11}p_{32}r
+t_{11}p_{31}r-
t_{11}p_{31}p_{32}r
\bigr\}
/
\bigl\{
\bigl[(1-x)y(p_{11}+p_{12}-p_{11}p_{12})-r\bigr](p_{31}p_{32}-p_{31}-p_{32})+1-r
\bigr\}$
\\
$\mathit{C}$ & $\mathcal{R}^\mathsf{cost}_{=?}[\mathsf{F\; succ \vee fail}]$ & 
expression (similar to the PMC expression for property $T$) not included for brevity, but available on project website
\\
\bottomrule
\end{tabular}
\end{table*}

\section{ePMC Theoretical Foundation \label{sect:theory}}

ePMC patterns are reoccurring ``fragments'' of parametric MCs with a single entry state and one or several output states. We formally define these concepts below. 

\begin{definition}\label{def:fragment}
A \emph{fragment} of a parametric Markov chain $M=(S,s_0,\mathbf{P},L)$ is a tuple 
$
  F=(Z,z_0,Z_\mathsf{out}),
$
where:
\squishlist
\item[-] $Z\subset S$ is a subset of transient MC states; 
\item[-] $z_0$ is the (only) \emph{entry state} of $F$, i.e., $\{z_0\}\!=\!\{z\!\in\! Z\mid \exists s\!\in\! S\!\setminus\! Z\:.\: \mathbf{P}(s,z)>0\}$;
\item[-] $Z_\mathsf{out} =\{z\in Z \mid \exists s\in S\setminus Z\:.\: \mathbf{P}(z,s)>0\}$ is the non-empty set of \emph{output states} of $F$, and all outgoing transitions from the output states are to states outside $Z$, i.e., $\mathbf{P}(z,z')=0$ for all $(z,z')\in Z_\mathsf{out}\times Z$.
\squishend
\end{definition}

\begin{example}
\label{ex:pmc-fragments}
The shaded areas of the parametric MC from Fig.~\ref{fig:running-pmc}a (each corresponding to an operation of the workflow from our running example) contain the three MC fragments:
\[
\begin{array}{l}
 F_1 = (\{s_0,s_1,s_2,s_3\}, s_0,\{s_2,s_3\})\\
 F_2 = (\{s_4,s_5,s_6,s_7,s_8\}, s_4, \{s_7,s_8\})\\
 F_3 = (\{s_9,s_{10},s_{11},s_{12}\}, s_9, \{s_{11},s_{12}\})
\end{array}
\]
As shown by this example, MC fragments may or may not contain cycles. 
\end{example}

\vspace*{1mm}
Given a fragment $F$ of a parametric MC $M$, ePMC performs parametric model checking by separately analysing two parametric MCs determined by $F$, and combining the results of the two analyses. As each of the two parametric MCs has fewer states and transitions than $M$, the overall result can be obtained in a fraction of the time required to analyse the original model $M$. 
The first of these parametric MCs is defined below.

\begin{definition}\label{def:associatedMC}
The \emph{Markov chain associated with a fragment} $F\!=\!(Z,z_0,Z_\mathsf{out})$ of a parametric MC $M\!=\!(S,s_0,\mathbf{P},L)$ is the Markov chain
$
  M_Z\!=\!(Z\cup\{e\},z_0,\mathbf{P}_Z,L_Z),
$
where $e$ is an additional, ``end'' state,  the transition probability matrix $\mathbf{P}_Z:(Z\cup\{e\})\times (Z\cup\{e\})\to [0,1]$ is given by 
\[
\mathbf{P}_Z(z,z') = \left\{
\begin{array}{ll}
1,& \textrm{if $z\in Z_\mathsf{out}\cup\{e\} \wedge z'=e$}\\
\mathbf{P}(z,z'),& \textrm{if $z\in Z\setminus Z_\mathsf{out} \wedge z'\neq e$}\\
0,&  \textrm{otherwise}
\end{array}
\right.,
\]
\mbox{and the atomic propositions for state $z\in Z$ are given by} 
\[
L_Z(z) = \left\{
\begin{array}{ll}
L(z)\cup\{\mathsf{z}\}, & \textrm{if $z\in Z_\mathsf{out}$}\\
\{\mathsf{e}\},& \textrm{if $z=e$}\\
L(z),& \textrm{otherwise}
\end{array}
\right. ,
\]
where $\mathsf{z}$ and $\mathsf{e}$ are atomic propositions that hold in state $z\in Z_\mathsf{out}$ and state $e$, respectively.
\end{definition}

\vspace*{1mm}
\begin{example}
Fig.~\ref{fig:running-pmc}b shows the parametric MCs associated with fragments $F_1$--$F_3$ from Example~\ref{ex:pmc-fragments}, obtained by:
\squishlist
\item[(i)] adding transitions of probability 1 from the output states in $Z_{\mathsf{out}_1}\!=\!\{s_2,s_3\}$, $Z_{\mathsf{out}_2}\!=\!\{s_7,s_8\}$ and $Z_{\mathsf{out}_3}\!=\!\{s_{11},s_{12}\}$ to additional states $e_1$, $e_2$ and $e_3$, respectively;
\item[(ii)] labelling the output states with the additional atomic propositions $\mathsf{s}_2$ and $\mathsf{s}_3$, $\mathsf{s}_7$ and $\mathsf{s}_8$, and $\mathsf{s}_{11}$ and $\mathsf{s}_{12}$, and the end states with the new atomic propositions $\mathsf{e}_1$ to $\mathsf{e}_3$.
\squishend
\end{example}

\vspace*{3mm}
The second parametric MC determined by a fragment $F$ and analysed by ePMC is obtained from the original MC by replacing all states from $F$ with a single state.

\begin{definition}
\label{def:abstractMC}
Given a fragment $F\!=\!(Z,z_0,Z_\mathsf{out})$ of a parametric Markov chain $M\!=\!(S,s_0,\mathbf{P},L)$, the \emph{abstract MC induced by  $F$} is 
$
  M'\!=\!(S',s'_0,\mathbf{P}',L'),
$
where:
\squishlist
\item[-] The state set $S'=(S\setminus Z) \cup \{\overline{z}\}$, where $\overline{z}$ is a new, \emph{abstract state} that stands for all the states from $Z$;
\item[-] The initial state $s'_0=s_0$, if $s_0$ is not the initial state of $Z$ (i.e., $z_0\neq s_0$), and $s_0'=\overline{z}$ otherwise;
\item[-] The transition probability  between states $s,s'\in S'$ is 
\[
\mathbf{P}'(s,s') \!=\! 
\left\{\!\!\!
\begin{array}{ll}
\mathbf{P}(s,s'), & \!\!\!\textrm{if $s\neq \overline{z} \wedge s'\neq \overline{z}$}\\
\mathbf{P}(s,z_0), & \!\!\!\textrm{if $s\neq \overline{z} \wedge s'=\overline{z}$}\\
\sum\limits_{z\in Z_\mathsf{out}}\!\! \mathit{prob}_z\mathbf{P}(z,s'), & \!\!\!\textrm{if $s=\overline{z}\wedge s'\neq\overline{z}$}\\
0, & \!\!\!\textrm{otherwise}
\end{array}
\right.
\]
where 
\begin{equation}
\label{eq:output-state-reachability}
\mathit{prob}_z=\mathcal{P}_{=?}[\mathrm{F}\: \mathsf{z}]
\end{equation}
is a reachability property calculated over the parametric Markov chain associated with the fragment $F$, for all output states $z\in Z_\mathsf{out}$;\footnote{Note that 
$\sum_{s'\in S'}\mathbf{P}'(\overline{z},s')=\sum_{s'\in S\setminus Z} \sum_{z\in Z_\mathsf{out}} \mathit{prob}_z\mathbf{P}(z,s')=$ 
\mbox{$\sum_{z\in Z_\mathsf{out}} \!\mathit{prob}_z\! \sum_{s'\in S\setminus Z} \!\mathbf{P}(z,s')\!=\!\sum_{z\in Z_\mathsf{out}} \!\mathit{prob}_z\!\cdot\! 1\!=\!\mathcal{P}_{=?}[\mathrm{F}\;\bigvee_{z\in Z_\mathsf{out}} \!z]$} $=1$ 
as required.}
\item[-] The labelling function $L'$ coincides with $L$ for the states from the original MC, and maps the new state $\overline{z}$ to the set of atomic propositions common to all states from $F$,
\[
L'(s)=\left\{
\begin{array}{ll}
L(s) & \textrm{if $s\in S\setminus Z$}\\
\bigcap\limits_{z\in Z} L(z) & \textrm{otherwise (i.e.\ if $s=\overline{z}$)}
\end{array}
\right..
\]
\squishend
Finally, for every structure of rewards $\mathit{rwd}$ defined over the Markov chain $M$, state $\overline{z}$ from the induced Markov chain is annotated with a reward 
\begin{equation}
\label{eq:fragment-reward}
\begin{array}{c}
  \overline{\mathit{rwd}}=\mathcal{R}^\mathsf{rwd}_{=?}\left[\mathrm{F}\;\mathsf{e}\right]
\end{array}
\end{equation}
calculated over the parametric MC $M_Z$ associated with $F$. Thus, $\overline{\mathit{rwd}}$ represents the cumulative reward to reach the end state of $M_Z$.
\end{definition}

\begin{example}
Consider again the parametric Markov chain from our running example (Fig.~\ref{fig:running-pmc}a). The corresponding abstract MC induced by fragment $F_1$ from Example~\ref{ex:pmc-fragments} is shown in Fig.~\ref{fig:running-pmc}c(i). This abstract MC is obtained by replacing all the states from $F_1$ with the single abstract state $\overline{z}_1$, and by using the rules from Definition~\ref{def:abstractMC} to find the outgoing transition probabilities and atomic propositions for $\overline{z}_1$. For example, the transition probability from $\overline{z}_1$ to $s_4$ is calculated as:
\[
\begin{split}
  \mathbf{P}'(\overline{z}_1,s_4) & = \sum\limits_{z\in \{s_2,s_3\}}\!\! \mathit{prob}_z\cdot\mathbf{P}(z,s_4)\\
  & = \mathit{prob}_{s_2}\cdot \mathbf{P}(s_2,s_4) + \mathit{prob}_{s_3}\cdot \mathbf{P}(s_3,s_4)\\ 
  & = \mathit{prob}_{s_2}\cdot x + \mathit{prob}_{s_3}\cdot 0\\
  & = \mathit{prob}_{s_2}\cdot x,
\end{split}
\]
where $\mathit{prob}_{s_2}\!\!=\mathcal{P}_{=?}[\mathrm{F}\: \mathsf{s}_2]$ and $\mathit{prob}_{s_3}\!\!=\mathcal{P}_{=?}[\mathrm{F}\: \mathsf{s}_3]=1-\mathit{prob}_{s_2}$ are reachability properties calculated over the parametric MC associated with fragment $F_1$ (cf.~Fig.~\ref{fig:running-pmc}b). As the two output-state reachability probabilities~(\ref{eq:output-state-reachability}) for fragment $F_1$ can be expressed in terms of a single probability, we use the notation $\mathit{prob}_1\!=\!\mathit{prob}_{s_2}$ for this probability in Fig.~\ref{fig:running-pmc}c(i). The transition probabilities from $\overline{z}_1$ to $s_9$ and $s_{13}$ are calculated similarly, and the transition probability from $s_{12}$ to $\overline{z}_1$ is simply $\mathbf{P}'(s_{12},\overline{z}_1)=\mathbf{P}(s_{12},s_0)=y$ (since the entry state of $F_1$ is $s_0$). All other transition probabilities from $\overline{z}_1$ to other states and from other states to $\overline{z}_1$ are zero. State $\overline{z}_1$ is labelled with the atomic proposition $\mathsf{op}_1$, which is the only label common to all states from the fragment $F_1$. Finally, $\overline{z}_1$ is annotated with the rewards $\mathit{time}_1\!=\!\mathcal{R}^\mathsf{RT}_{=?}[\mathrm{F} \;s_2\vee s_3]$ and $\mathit{cost}_1\!=\!\mathcal{R}^\mathsf{cost}_{=?}[\mathrm{F} \;s_2\vee s_3]$ computed over the parametric MC associated with $F_1$.

Fig.~\ref{fig:running-pmc}c(ii) shows the abstract Markov chain obtained after all three fragments $F_1$--$F_3$ from Example~\ref{ex:pmc-fragments} were used to simplify the initial MC from Fig.~\ref{fig:running-pmc}a. Note how even for the small MC from our running example, the abstract MC from Fig.~\ref{fig:running-pmc}c(ii) is much simpler than the initial MC from Fig.~\ref{fig:running-pmc}a; the abstract MC has only 5~states and 10  transitions, compared to 15 states and 25 transitions for the initial MC.
\end{example}

\vspace*{2mm}
The ePMC computation of unbounded until properties $\mathcal{P}_{=?}[\Phi_1\,\mathrm{U}\,\Phi_2]$ (and thus also of reachability properties $\mathcal{P}_{=?}[\mathrm{F}\,\Phi] = \mathcal{P}_{=?}[\mathit{true}\,\mathrm{U}\,\Phi]$) is underpinned by the following result, whose proof is provided in Appendix~A. 

\begin{theorem}
\label{th:until}
Let $F$ be a fragment of a parametric Markov chain $M$, and $\Phi_1$ and $\Phi_2$ two PCTL state formulae over $M$. If every atomic proposition in $\Phi_1$ and $\Phi_2$ either holds in all states from $F$ or holds in no state from $F$, then the PMC of the until PCTL formula $\mathcal{P}_{=?}[\Phi_1\,\mathrm{U}\,\Phi_2]$ over $M$ yields an expression equivalent to that produced by the PMC of the formula over the abstract Markov chain induced by $F$. 
\end{theorem}

The repeated application of Theorem~\ref{th:until} reduces the computation of until properties $\mathcal{P}_{=?}[\Phi_1\,\mathrm{U}\,\Phi_2]$ of a parametric MC with multiple fragments $F_1$, $F_2$, \ldots to computing:
\squishlist
\item[1)] the output-state reachability probabilities~(\ref{eq:output-state-reachability}) for the parametric MCs associated with $F_1$, $F_2$, \ldots;
\item[2)] $\mathcal{P}_{=?}[\Phi_1\,\mathrm{U}\,\Phi_2]$ for the parametric MC induced by the fragments 
\squishend
and combining the results from the two ePMC stages into a set of algebraic formulae over the parameters of the original MC. The parametric MCs from these stages are typically much simpler than the original, ``monolithic'' MC, and much faster to analyse. In addition, ePMC focuses on frequently used domain-specific fragments $F_1$, $F_2$, \ldots, and thus stage~1 only needs to be executed once for a domain. Note that a result similar to Theorem~\ref{th:until} is not available for bounded until properties $\mathcal{P}_{=?}[\Phi_1\,\mathrm{U}^{\leq k}\,\Phi_2]$ because the abstract MC induced by a set of fragments $F_1,F_2,\ldots$ does not preserve the path lengths from the original MCs. 

\begin{example}
\label{ex_P1_P2}
We use the above two-stage method to compute properties $P_1$ and $P_2$ from our running example (cf.~Table~\ref{tab:running-example-properties}). 

In stage~1 we compute the output-state reachability properties~(\ref{eq:output-state-reachability}) for the parametric MCs associated with fragments $F_1$--$F_3$ (cf.~Fig.~\ref{fig:running-pmc}b): 
\squishlist
\item for the MC associated with $F_1$, $\mathit{prob}_1\!=p_{11}+(1-p_{11})p_{12}$;  
\item for the MC associated with $F_2$, $\mathit{prob}_2=\alpha_1p_{21}+\alpha_2p_{22}$;  
\item for the MC associated with $F_3$, $\mathit{prob}_3\!=(p_{31}+(1\!-p_{31})$ $p_{32})/\left(1-(1\!-\!p_{31})(1\!-\!p_{32})r\right)$. 
\squishend
We computed these algebraic expressions manually, based on the MCs from Fig.~\ref{fig:running-pmc}b. However, they can also be obtained using one of the model checkers mentioned earlier (i.e., PARAM, PRISM and Storm), or can be taken directly from our ePMC repository of such expressions for the service-based systems domain (see Section~\ref{sect:sbs} later in the paper). 

In stage~2, we use a probabilistic model checker (we used Storm) to compute $P_1$ and $P_2$ over the induced parametric MC from Fig.~\ref{fig:running-pmc}c.
The \colorbox{gray!30}{shaded formulae} from Table~\ref{tab:running-example-properties2} show the expressions obtained for $P_1$ and $P_2$,  preceded by the results from the first ePMC stage. 

Expectedly, the set of formulae from Table~\ref{tab:running-example-properties2} 
is much simpler than the ``monolithic'' $P_1$ and $P_2$ formulae from Table~\ref{tab:running-example-properties}. As we will show in Section~\ref{sect:evaluation}, this difference is even more significant for larger models, making the computation and evaluation of ``monolithic'' formulae challenging for existing PMC techniques.
\end{example}

\begin{table}
\caption{ePMC of the QoS properties from the running example}
\label{tab:running-example-properties2}
\centering
\renewcommand{\arraystretch}{1.5}

\vspace*{-1mm}
\begin{tabular}{l} 
\toprule
\textbf{Output-state reachability formulae computed in stage 1 of ePMC}\\
\midrule 
$
\begin{array}{l}
\textrm{\colorbox{gray!30}{$\mathit{prob}_1=p_{11}+(1-p_{11})p_{12}\qquad\quad\;\!\mathit{prob}_2=\alpha_1p_{21}\!+\!\alpha_2p_{22}\qquad\;\!$}}\\[-0.1mm]
\textrm{\colorbox{gray!30}{$\mathit{prob}_3=\frac{p_{31}+(1-p_{31})p_{32}}{1-(1-p_{31})(1-p_{32})r}\textrm{\hspace*{4.38cm}}$}}\\
\textrm{\colorbox{white}{$\mathit{time}_1=t_{11}+(1-p_{11})t_{12}\qquad\quad\;\mathit{cost}_1=c_{11}+(1-p_{11})c_{12}$}}\\
\textrm{\colorbox{white}{$\mathit{time}_2=\alpha_1t_{21}\!+\!\alpha_2t_{22}\qquad\qquad\quad\,\,\mathit{cost}_2=\alpha_1c_{21}\!+\!\alpha_2c_{22}$}}\\
\textrm{\colorbox{white}{$\mathit{time}_3=\frac{t_{31}+(1-p_{31})t_{32}}{1-(1-p_{31})(1-p_{32})r}\qquad\quad\!\mathit{cost}_3=\frac{c_{31}+(1-p_{31})c_{32}}{1-(1-p_{31})(1-p_{32})r}$}}\\[1.5mm] 
\end{array}
$\\ 
\midrule
\midrule
\multicolumn{1}{c}{\textbf{Property-specific formulae computed in stage 2 of ePMC}}\\
\end{tabular}
\begin{tabular}{p{0.55cm}p{2.25cm}p{4.8cm}} 
\midrule
\textbf{Prop.} & \textbf{PCTL formula} & \hspace*{2mm}\textbf{ePMC set of formulae}\\
\midrule
$
\begin{array}{l}
P_1\\[1.1mm]
P_2\\[1.1mm]
\mathit{T}\\[1.1mm]
\mathit{C}
\end{array}
$ 
& 
$
\begin{array}{l}
\mathcal{P}_{=?}[\mathsf{F\; succ}]\\[1.1mm]
\mathcal{P}_{=?}[\neg \mathsf{op}_3\; \mathrm{U}\; \mathsf{fail}]\\[1.1mm]
\mathcal{R}^\mathsf{time}_{=?}[\mathsf{F\; succ \vee fail}]\\[1.1mm]
\mathcal{R}^\mathsf{cost}_{=?}[\mathsf{F\; succ \vee fail}]
\end{array}
$ 
& 
$
\begin{array}{l}
\textrm{\colorbox{gray!30}{$P_1=\frac{\mathit{prob}_1(x\mathit{prob}_2+(1-x)(1-y)\mathit{prob}_3)}{1-(1-x)y\mathit{prob}_1\mathit{prob}_3}\;$}}\\[-0.3mm]
\textrm{\colorbox{gray!30}{$P_2=1-\mathit{prob}_1+x\mathit{prob}_1(1-\mathit{prob}_2)\!$}}\\
\textrm{\colorbox{white}{$\mathit{T}=\frac{\mathit{time}_1 + \mathit{prob}_1(x\mathit{time}_2 + (1-x)\mathit{time}_3)} {1-(1-x)y\mathit{prob}_1\mathit{prob}_3}$}}\\
\textrm{\colorbox{white}{$\mathit{C}=\frac{\mathit{cost}_1 + \mathit{prob}_1(x\mathit{cost}_2 + (1-x)\mathit{cost}_3)} {1-(1-x)y\mathit{prob}_1\mathit{prob}_3}$}}
\end{array}
$\\
\bottomrule
\end{tabular}

\vspace*{-2mm}
\end{table}

\vspace*{1mm}
The final result from this section allows the efficient parametric model checking of reachability reward properties.

\begin{theorem}
\label{th:rewards}
Let $F$ be a fragment of a parametric Markov chain $M$, and $T$ a set of states from $M$. If $T$ includes no state from $F$, then the PMC of the PCTL formula 
$\mathcal{R}_{=?}[\mathrm{F}\; T]$ (i.e., the cumulative reward to reach a state from $T$) over $M$ yields an expression equivalent to that produced by the PMC of the formula over the abstract MC $M'$ induced by $F$. 
\end{theorem}

Analogous to Theorem~\ref{th:until}, this result (which we prove in Appendix~A) reduces the computation of cumulative reachability reward properties $\mathcal{R}_{=?}[\mathrm{F}\; T]$ of a parametric MC with  fragments $F_1$, $F_2$, \ldots to computing:
\squishlist
\item[1)] the per-fragment cumulative reachability reward properties~(\ref{eq:fragment-reward}) for the MCs associated with $F_1$, $F_2$, \ldots;
\item[2)] $\mathcal{R}_{=?}[\mathrm{F}\; T]$ for the MC induced by these fragments
\squishend
and combining the results from the two stages into a set of algebraic formulae over the parameters of the original MC.  
Note that results similar to Theorem~\ref{th:rewards} are not available for instantaneous, cumulative and steady-state rewards formulae because the abstract MC induced by $F_1$, $F_2$, \ldots does not preserve the path lengths and the rewards structures of the original MC.

\begin{example}
We use ePMC to calculate properties $\mathit{time}$ and $\mathit{cost}$ from our running example (cf.~Table~\ref{tab:running-example-properties}), starting with the cumulative reachability reward properties~(\ref{eq:fragment-reward}) for fragments $F_1$--$F_3$, i.e., $\overline{t}_i$ and $\overline{c}_i$ for $i=1,2,3$. The resulting formulae, which we obtained manually (but which can also be obtained using a PMC tool) are shown in the top half of Table~\ref{tab:running-example-properties2}. For the second ePMC stage, we used the model checker Storm to obtain the algebraic expressions for $\mathit{time}$ and $\mathit{cost}$ from the lower half of Table~\ref{tab:running-example-properties2}. 

As in Example~\ref{ex_P1_P2}, ePMC produced a set of formulae that is far simpler than the ``monolithic'' $\mathit{time}$ and $\mathit{cost}$ from Table~\ref{tab:running-example-properties}. Note that we do not compare the analysis time of our ePMC method with that of existing PMC  here or in Example~\ref{ex_P1_P2} because for the simple system from our running example the two analysis times are similar. However, we do provide an extensive comparison of these analysis times for larger systems in our evaluation of ePMC from Section~\ref{sect:evaluation}.
\end{example}

\section{Implementation \label{sect:tool}}

We developed a \emph{pattern-aware parametric model checker} that implements the theoretical results from the previous section. This tool, which is freely available on our project website \url{https://www.cs.york.ac.uk/tasp/ePMC}, automates the second stage of ePMC. As shown in Fig.~\ref{fig:method}, the ePMC tool uses a domain-specific repository of QoS-property expressions to analyse PCTL-specified QoS properties of a parametric Markov chain annotated with pattern instances. 

The domain-specific repository comprises entries with the general format:

\begin{small}
\begin{lstlisting}
 pattern_name(parameter_list):
  property_name=expr, ..., property_name=expr;
\end{lstlisting}
\end{small}

\noindent
Each such entry defines algebraic expressions for the reachability properties~(\ref{eq:output-state-reachability}) and the reachability reward properties~(\ref{eq:fragment-reward}) of a parametric MC fragment commonly used within the domain of interest, i.e., a \emph{modelling pattern}. 

\begin{example}
Table~\ref{tab:SBS-repository-expressions} shows a part of the ePMC repository for the service-based systems domain. This part includes the three patterns used by the operations from our running example (\textsf{SEQ} for $\mathsf{op}_1$, \textsf{PROB} for $\mathsf{op}_2$, and \textsf{SEQ\_R} for $\mathsf{op}_3$), such that the formulae from the top half of Table~\ref{tab:running-example-properties2} can be obtained (without any calculations) by instantiating the relevant patterns:
\begin{equation}
\label{eq:pattern_instantiations}
\!\begin{array}{lll}
  \mathit{prob}_1&\!\!\!\!\!=\!\!\!\!\!&\mathsf{SEQ}(p_{11},c_{11},t_{11},p_{12},c_{12},t_{12}).\mathsf{prob}\\
  \mathit{prob}_2&\!\!\!\!\!=\!\!\!\!\!&\mathsf{PROB}(\alpha_1,p_{21},c_{21},t_{21},\alpha_2,p_{22},c_{22},t_{22}).\mathsf{prob}\\
  \mathit{prob}_3&\!\!\!\!\!=\!\!\!\!\!&\mathsf{SEQ\_R}(p_{31},c_{31},t_{31},p_{32},c_{32},t_{32},r).\mathsf{prob}\\
  \mathit{time}_1&\!\!\!\!\!\!=\!\!\!\!\!&\mathsf{SEQ}(p_{11},c_{11},t_{11},p_{12},c_{12},t_{12}).\mathsf{time}\\
  \mathit{cost}_1&\!\!\!\!\!\!=\!\!\!\!\!&\mathsf{SEQ}(p_{11},c_{11},t_{11},p_{12},c_{12},t_{12}).\mathsf{cost}\\
  \mathit{time}_2&\!\!\!\!\!\!=\!\!\!\!\!&\mathsf{PROB}(\alpha_1,p_{21},c_{21},t_{21},\alpha_2,p_{22},c_{22},t_{22}).\mathsf{time}\\
  \mathit{cost}_2&\!\!\!\!\!\!=\!\!\!\!\!&\mathsf{PROB}(\alpha_1,p_{21},c_{21},t_{21},\alpha_2,p_{22},c_{22},t_{22}).\mathsf{cost}\\
  \mathit{time}_3&\!\!\!\!\!\!=\!\!\!\!\!&\mathsf{SEQ\_R}(p_{31},c_{31},t_{31},p_{32},c_{32},t_{32},r).\mathsf{time}\\
  \mathit{cost}_3&\!\!\!\!\!\!=\!\!\!\!\!&\mathsf{SEQ\_R}(p_{31},c_{31},t_{31},p_{32},c_{32},t_{32},r).\mathsf{cost}
\end{array}
\end{equation}
\end{example}

\begin{table}
\caption{Fragment of the ePMC repository of QoS-property expressions for the service-based systems domain, comprising expressions for the probability of pattern invocation success \texttt{prob}, expected \texttt{cost}, and expected execution \texttt{time}}
\label{tab:SBS-repository-expressions}

\vspace*{-4mm}
\hrulefill

\vspace*{-1mm}
\begin{lstlisting}
SEQ(p1,c1,t1,p2,c2,t2): 
 prob=p1+(1-p1)*p2, cost=c1+c2*(1-p1), 
 time=t1+(1-p1)*t2;
...
PROB(x1,p1,c1,t1,x2,p2,c2,t2):
 prob=x1*p1+x2*p2, cost=x1*c1+x2*c2,
 time=x1*t1+x2*t2;
...
SEQ_R(p1,c1,t1,p2,c2,t2,r):
 prob=(p1+(1-p1)*p2)/(1-(1-p1)*(1-p2)*r),
 cost=(c1+(1-p1)*c2)/(1-(1-p1)*(1-p2)*r),
 time=(t1+(1-p1)*t2)/(1-(1-p1)*(1-p2)*r);
...
\end{lstlisting}

\vspace*{-3mm}
\hrulefill
\end{table}

\vspace*{3mm}
The ePMC model checker supports the analysis of pattern-annotated parametric Markov chains specified in the high-level modelling language used by PRISM and Storm. This language models a system as the parallel composition of a set of \emph{modules}. The state of a \emph{module} is encoded by a set of finite-range local variables, and its state transitions are defined by probabilistic guarded commands that change these variables, and have the general form:\\[-2mm]

 [$\mathit{action}$] $\mathit{guard}$ $\rightarrow$ $e_1$:$\,\mathit{update}_1$ + \ldots + $e_N$:$\,\mathit{update}_N$;\\[-2mm]

\noindent
In this command, $\mathit{guard}$ is a boolean expression over all model variables. If $\mathit{guard}$ evaluates to $\mathit{true}$, the arithmetic expression $e_i$, $1\leq i\leq N$, gives the probability with which the $\mathit{update}_i$ change of the module variables occurs. When the label $\mathit{action}$ is present, all modules comprising commands with this $\mathit{action}$ have to synchronise (i.e., can only carry out one of these commands simultaneously). 

\begin{figure}
\centering
\includegraphics[width=\hsize]{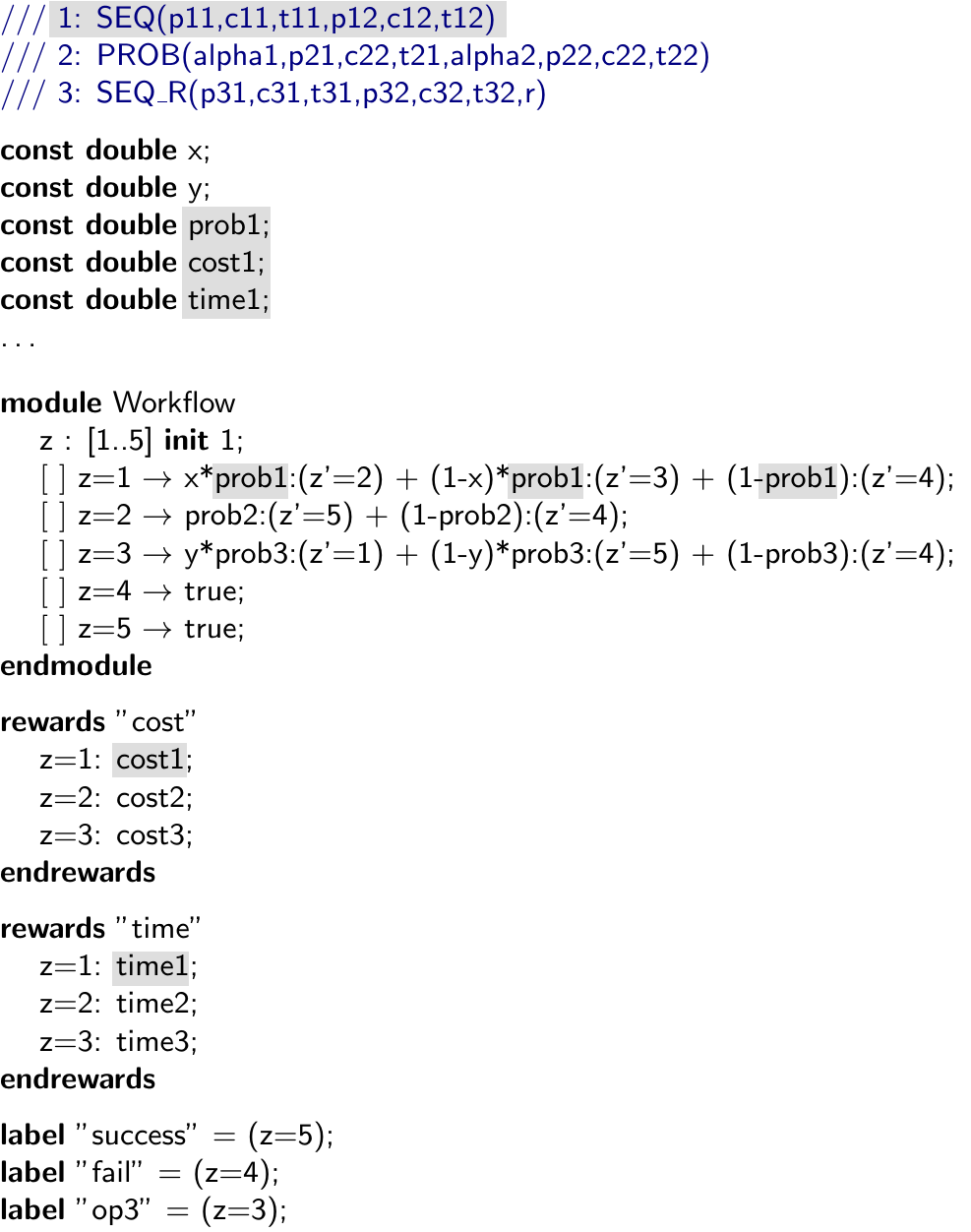}

\vspace*{-1.5mm}
\caption{Pattern-annotated parametric Markov chain for the service-based system from the running example \label{fig:annotated-SBS-model}}

\vspace*{-3mm}
\end{figure}

\begin{example}
Fig.~\ref{fig:annotated-SBS-model} shows how the parametric MC from Fig.~\ref{fig:running-pmc}c(ii) and its reward structures and labels are specified in this high-level modelling language. The values $\mathsf{z}\!=\!1$ to $\mathsf{z}\!=\!5$ of the local variable \textsf{z} from the \textsf{Workflow} module correspond to states $\overline{z}_1$, $\overline{z}_2$, $\overline{z}_3$, $s_{13}$ and $s_{14}$ from Fig.~\ref{fig:running-pmc}c(ii), respectively. The model parameters  
$\mathsf{prob1}$ to $\mathsf{prob3}$,  
$\mathsf{cost1}$ to $\mathsf{cost3}$ and  
$\mathsf{time1}$ to $\mathsf{time3}$ 
are associated with the pattern annotations from the lines starting with a triple forward slash `$\mathsf{///}$'. 
These annotations 
tell the ePMC model checker that the model has parameters associated with QoS properties of modelling patterns from the repository in Table~\ref{tab:SBS-repository-expressions}. For example, the shaded pattern annotation 
at the top of Fig.~\ref{fig:annotated-SBS-model} specifies that the QoS properties $\mathsf{prob}$, $\mathsf{cost}$ and $\mathsf{time}$ of the $\mathsf{SEQ}$ modelling pattern appear as parameters named $\mathsf{prob1}$, $\mathsf{cost1}$ and $\mathsf{time1}$ in the model, where the id `$1$' is provided before the pattern name. The occurrences of these parameters are also shaded in Fig.~\ref{fig:annotated-SBS-model}. 
\end{example}

The general format of an ePMC pattern annotation is:

\begin{small}
\begin{lstlisting}
 /// id: pattern_name(actual_parameter_list)
\end{lstlisting}
\end{small}

\begin{table*}
\caption{Modelling patterns for the implementation of SBS operations using $n$ functionally-equivalent services}
\label{tab:sbs_patterns}
\centering

\vspace*{-1.5mm}
\begin{tabular}{p{5.2cm}p{11.8cm}} 
\toprule
\textbf{Pattern} & \textbf{Description}\\
\midrule 
$\mathsf{SEQ}(p_1,c_1,t_1,\ldots,p_n,c_n,t_n)$ &The $n$ services are invoked in order, stopping after the first successful invocation or after the last service.\\[1mm]
$\mathsf{PAR}(p_1,c_1,t_1,\ldots,p_n,c_n,t_n)^\dagger$ &The $n$ services are all invoked at the same time, and the operation uses the first result returned by a successful invocation (if any).\\[1mm]
$\mathsf{PROB}(x_1,p_1,c_1,t_1,\ldots,x_n,p_n,c_n,t_n)$ & A single service is invoked; $x_i$ gives the probability that this is service $i$, where $\sum_{i=1}^n x_i=1$.\\[1mm]
\textbf{$\mathsf{SEQ\_R}(p_1,c_1,t_1,\ldots,p_n,c_n,t_n,r)$} & The $n$ services are invoked in order as for the $\mathsf{SEQ}$ pattern; if all $n$ invocations fail, the execution of the operation is retried (from service $1$) with probability $r$ or the operation fails with probability $1-r$. \\[1mm]
$\mathsf{SEQ\_R1}(p_1,c_1,t_1,r_1,\ldots,p_n,c_n,t_n,r_n)$ & The services are invoked in order. If service $i$ fails, it is reinvoked with probability $r_i$; with probability $1-r_i$, the operation is attempted using service $i+1$ (if $i<n$) or fails (if $i=n$).\\[1mm]
\textbf{$\mathsf{PAR\_R}(p_1,c_1,t_1,\ldots,p_n,c_n,t_n,r)^\dagger$} & All $n$ services are invoked as for the $\mathsf{PAR}$ pattern; if all $n$ invocations fail, the execution of the operation is retried with probability $r$ or the operation fails with probability $1-r$.\\[1mm]
\textbf{$\mathsf{PROB\_R}(x_1,p_1,c_1,t_1,\ldots,$ $\textsf{\hspace*{30mm}}  x_n,p_n,c_n,t_n,r)$} & Like for $\mathsf{PROB}$, a single service $i$ is invoked; if the invocation fails, the $\mathsf{PROB}$ pattern is retried with probability $r$ or the operation fails with probability $1-r$.\\[1mm]
\textbf{$\mathsf{PROB\_R1}(x_1,p_1,c_1,t_1,r_1,\ldots,$ $\textsf{\hspace*{30mm}} x_n,p_n,c_n,t_n,r_n)$} &Like for $\mathsf{PROB}$, a single service $i$ is invoked; however, its invocation is retried after failure(s) with probability $r_i$ or the operation fails with probability $1-r_i$.\\
\bottomrule
\multicolumn{2}{l}{$^\dagger$Pattern unsuitable for non-idempotent operations (e.g.\ credit card payment in an e-commerce SBS)}
\end{tabular}

\vspace*{-2mm}
\end{table*}

\noindent
This annotation indicates that some or all QoS properties of the pattern appear as parameters in the parametric MC, with the name of each such parameter obtained by appending the $\mathsf{id}$ from the annotation to the name of the QoS property.

Given a domain-specific repository of QoS-property expressions, a pattern-annotated parametric MC and a set of PCTL-encoded QoS properties, the ePMC model checker uses the theoretical results from Section~\ref{sect:theory} to compute a set of algebraic formulae comprising:
\squishlist
\item[1.] Formulae for every MC parameter associated with a modelling pattern listed in the model annotations. These formulae are obtained by instantiating the expressions from the repository, as shown in~(\ref{eq:pattern_instantiations}). The top half of Table~\ref{tab:running-example-properties2} shows these formulae for the MC in Fig.~\ref{fig:annotated-SBS-model}.
\item[2.] A formula for each analysed QoS property, obtained by applying standard parametric model checking, i.e., by ignoring the pattern annotations of the parametric MC. The bottom half of Table~\ref{tab:running-example-properties2} shows these formulae for the four QoS properties from our running example.
\squishend
The ePMC tool can be configured to use PRISM or Storm for the computation of the latter formulae, and outputs the combined set of algebraic formulae as a MATLAB file, ready for evaluation or further analysis with MATLAB.

\section{ePMC of Service-Based Systems \label{sect:sbs}}

Service-based systems (SBSs) enable the effective development of new applications through the integration of third-party and in-house components implemented as services. SBSs are widely used, including in business-critical applications from e-commerce, online banking and e-government, and evolve frequently as a result of maintenance or self-adaptation. This evolution often requires the QoS analysis of alternative SBS implementations which deliver the same functionality, to select an implementation that meets the QoS requirements of the system. The alternative SBS implementations differ in the way in which they use the multiple functionally-equivalent services that are available for each of their operations. Given $n\geq 1$ services that can perform the same SBS operation with probabilities of success $p_1$, $p_2$, \ldots, $p_n$, costs $c_1$, $c_2$, \ldots, $c_n$, and execution times $t_1$, $t_2$, \ldots, $t_n$, the operation can be implemented using one of the patterns described in Table~\ref{tab:sbs_patterns}. As we discuss further in Section~\ref{sect:related}, variants of the first three patterns have been widely used in related research (e.g. in \cite{5611553,CARDOSO2004,zeng2004qos}), while---to the best of our knowledge---the remaining patterns from Table~\ref{tab:sbs_patterns} have not been considered before.

As indicated earlier in the paper, our ePMC method is well suited for the SBS domain, as the operation implementation patterns from Table~\ref{tab:sbs_patterns} correspond to component modelling patterns whose reliability, cost and execution time can be obtained in the first stage of the method. Table~\ref{tab:sbs_expressions} and the following theorem (which we prove in Appendix~A) provide the repository of (manually derived) closed-form expressions for all patterns from Table~\ref{tab:sbs_patterns} and three key QoS properties of SBS operations. 

\begin{table*}
\caption{Complete ePMC repository of QoS-property expressions for the SBS domain}
\label{tab:sbs_expressions}

\vspace*{-1.5mm}
\centering
\begin{tabular}{p{1.3cm}p{4cm}p{5.5cm}p{5.3cm}} 
\toprule
\textbf{Pattern} & \textbf{Success probability} & \textbf{Expected cost} & \textbf{Expected execution time} \\
\midrule 
$\mathsf{SEQ}$ & $p_\mathsf{SEQ}=1-\prod_{i=1}^n (1-p_i)$ & $c_\mathsf{SEQ}=c_1+\sum_{i=2}^n \left(\prod_{j=1}^{i-1}(1-p_j)\right)c_i$ & $t_\mathsf{SEQ}=t_1+\sum_{i=2}^n \left(\prod_{j=1}^{i-1}(1-p_j)\right)t_i$\\[1mm]
$\mathsf{PAR}^\dagger$ & $p_\mathsf{PAR}=p_\mathsf{SEQ}$ & $c_\mathsf{PAR}=\sum_{i=1}^n c_i$ & $t_\mathsf{PAR}=\widetilde{p}_1t_1+\sum_{i=2}^n \left(\prod_{j=1}^{i-1}(1-p_j)\right)\widetilde{p}_it_i$\\[1mm]
$\mathsf{PROB}$ & $p_\mathsf{PROB}=\sum_{i=1}^n x_ip_i$ & $c_\mathsf{PROB}=\sum_{i=1}^n x_ic_i$ & $t_\mathsf{PROB}=\sum_{i=1}^n x_it_i$\\[1mm]
$\mathsf{SEQ\_R}$ & $p_\mathsf{SEQ\_R}=\frac{p_\mathsf{SEQ}}{1-(1-p_\mathsf{SEQ})r}$ & $c_\mathsf{SEQ\_R}=\frac{c_\mathsf{SEQ}}{1-(1-p_\mathsf{SEQ})r}$ & $t_\mathsf{SEQ\_R}=\frac{t_\mathsf{SEQ}}{1-(1-p_\mathsf{SEQ})r}$\\[2mm]
$\mathsf{SEQ\_R1}^\ddagger$ & $p_\mathsf{SEQ\_R1}=1-\prod_{i=1}^n (1-p'_i)$ & $c_\mathsf{SEQ\_R1}=c'_1+\sum_{i=2}^n \left(\prod_{j=1}^{i-1}(1-p'_j)\right)c'_i$ & $t_\mathsf{SEQ\_R1}=t'_1+\sum_{i=2}^n \left(\prod_{j=1}^{i-1}(1-p'_j)\right)t'_i$\\[1mm]
$\mathsf{PAR\_R}$ & $p_\mathsf{PAR\_R}=p_\mathsf{SEQ\_R}$ & $c_\mathsf{PAR\_R}=\frac{c_\mathsf{PAR}}{1-(1-p_\mathsf{PAR})r}$ & $t_\mathsf{PAR\_R}=\frac{t_\mathsf{PAR}}{1-(1-p_\mathsf{PAR})r}$ \\[2mm]
$\mathsf{PROB\_R}$ & $p_\mathsf{PROB\_R}=\frac{p_\mathsf{PROB}}{1-(1-p_\mathsf{PROB})r}$ & $c_\mathsf{PROB\_R}=\frac{c_\mathsf{PROB}}{1-(1-p_\mathsf{PROB})r}$ & $t_\mathsf{PROB\_R}=\frac{t_\mathsf{PROB}}{1-(1-p_\mathsf{PROB})r}$\\
$\mathsf{PROB\_R1}^\ddagger$ & $p_\mathsf{PROB\_R1}=\sum_{i=1}^n x_ip'_i$ & $c_\mathsf{PROB\_R1}=\sum_{i=1}^n x_ic'_i$ & $t_\mathsf{PROB\_R1}=\sum_{i=1}^n x_it'_i$ \\[2mm]
\bottomrule
\multicolumn{4}{l}{$^\dagger$assuming that the $n$ services are ordered such that $t_1\leq t_2\leq\cdots\leq t_n$, with $\widetilde{p}_i=p_i$ for $i<n$ and $\widetilde{p}_n=1$}\\
\multicolumn{4}{l}{$^\ddagger$ $p'_i=\frac{p_i}{1-(1-p_i)r_i}$, $c'_i=\frac{c_i}{1-(1-p_i)r_i}$ and $t'_i=\frac{t_i}{1-(1-p_i)r_i}$ for all $i=1,2,\ldots,n$}
\end{tabular}

\vspace*{-2mm}
\end{table*}

\begin{theorem}
The closed-form expressions from Table~\ref{tab:sbs_expressions} specify the success probability, the expected cost, and the expected execution time for each of the SBS-operation implementation patterns from Table~\ref{tab:sbs_patterns}. 
\end{theorem}

\noindent
As we show experimentally in Section~\ref{sect:evaluation}, ePMC can use the repository of closed-form expressions from Table~\ref{tab:sbs_expressions} to efficiently compute reliability, cost and response-time QoS properties of realistic SBS designs that leading model checkers take a very long time to verify, or cannot handle at all due to out-of-memory or timeout errors. Furthermore, as also shown in Section~\ref{sect:evaluation}, our method yields closed-form expressions that are more compact and take far less time to evaluate than the expressions produced by traditional PMC.

\section{ePMC of Multi-Tier Architectures \label{sect:multitier}}

As a second application domain for ePMC, we consider the deployment of software systems with a multi-tier architecture on a set of servers. For improved reliability and throughput, these systems often use \emph{horizonal distribution} within some or all tiers, i.e.\ they have instances of these tiers running on multiple servers. In this section, we devise a repository of \emph{server modelling patterns} for analysing reliability properties of such systems. To this end, we consider different types of server on which $n_1\!\geq\! 1$, $n_2\!\geq\! 1$, \ldots, $n_m\!\geq\! 1$ instances of $m$ of the system tiers were deployed.

Our (non-exhaustive) set of server modelling patterns is presented in Table~\ref{tab:server_patterns}. The $\mathsf{BASIC}$ pattern from this table corresponds to a server whose failure leads to the immediate loss of all tier instances running on the server, while the $\mathsf{VIRTUALIZED}$ and $\mathsf{VIRTUALIZED\textrm{-}M}$(onitored) patterns correspond to servers where each instance of a tier is running within a separate virtual machine (VM). The difference between the two types of virtualized server is that the second type has a \emph{monitor} component capable of detecting imminent server failures early enough to allow the migration of the VMs to other servers.

\begin{table*}
\caption{Server modelling patterns for the reliability analysis of multi-tier architecture deployments}
\label{tab:server_patterns}
\centering

\vspace*{-0.5mm}\begin{tabular}{p{5.2cm}p{11.8cm}} 
\toprule
\textbf{Pattern} & \textbf{Description}\\
\midrule 
$\mathsf{BASIC}(n_1,n_2,\ldots, n_m,p)$ & The $n_1,n_2,\ldots, n_m$ tier instances are deployed on a server whose probability of remaining operational throughout a time period of interest (e.g.\ a month) is $p$; if the server fails, all tier instances are lost.\\[1mm]
$\mathsf{VIRTUALIZED}(n_1,n_2,\ldots, n_m,p,p_\mathsf{VM})$ & Each of the $n_1,n_2,\ldots, n_m$ tier instances runs within its own virtual machine (VM) on a server whose probability of remaining operational during the time period of interest is $p$; additionally, each VM has a probability $p_\mathsf{VM}$ of remaining operational during the same time period, independently of the other VMs.\\[1mm]
$\mathsf{VIRTUALIZED\textrm{-}M}(n_1,n_2,\ldots, n_m,p,$ 
$\textsf{\hspace*{24mm}}p_\mathsf{detect},p_\mathsf{migrate},r,p_\mathsf{VM})$ & The $n_1,n_2,\ldots, n_m$ tier instances are deployed within different VMs (each with probability $p_\mathsf{VM}$ of not failing), on a server whose probability of remaining operational is $p$. If the server does fail, there is a probability $p_\mathsf{detect}$ that a software monitor will detect the approaching failure before it happens, allowing the VMs to be migrated to other servers. The migration of each VM succeeds with probability $p_\mathsf{migrate}$, and is retried with probability $r$ in case of failure.\\
\bottomrule
\end{tabular}
\end{table*}

We assume that the engineers responsible for deploying a multi-tier software system on a combination of such servers need to assess the following reliability properties of alternative deployment options:
\squishlist
\item[1)] The probability $P_\mathsf{FAIL}$ of system failure due to all instances of a tier failing within a time period of interest;
\item[2)] The probability $P_\mathsf{SPF}$ of a single point of failure (i.e.\ a tier with a single operational instance) occurring within the analysed time period.
\squishend
ePMC can support the analysis of these properties by using a repository of QoS-property expressions comprising entries for each probability $p_{b_1,b_2,\ldots,b_m}$ that $b_i\!\in\!\{0, 1, 2+\}$ instances of tier $i$, $i\in\{1,2,\dots,m\}$, remain operational on the types of server from Table~\ref{tab:server_patterns} at the end of the analyzed time period. For example, $p_{0,2+}$ represents the probability that a server running instances of two tiers at the beginning of the analysed period is left with no instance of the first tier and with two or more instances of the second tier at the end of the period. 
Although $3^{m}$ expressions need to be computed for these probabilities and each type of server, this is feasible because $m$ is a small number (e.g., $m\leq 3$ for a three-tier architecture).

Table~\ref{tab:server_expressions} and the following theorem (which we prove in Appendix~A) provide the repository of (manually derived) closed-form expressions for all server modelling patterns from Table~\ref{tab:server_patterns}.

\begin{table*}
\caption{ePMC repository of QoS-property expressions for the multi-tier architecture domain}
\label{tab:server_expressions}

\vspace*{-0.5mm}
\centering
\begin{tabular}{p{2.4cm}p{14.6cm}} 
\toprule
\textbf{Pattern} & \textbf{Probability that $b_i\!\in\!\{0, 1, 2+\}$ tier-$i$ instances, $i\in\{1,2,\dots,m\}$, remain operational} \\
\midrule 
$\mathsf{BASIC}$ & 
$p_{b_1,b_2,\ldots,b_m}=\left\{ \begin{array}{ll} 
p, & \textrm{if } \forall i\in\{1,2,\ldots,m\}.(n_i>1\wedge b_i=2+)\vee(n_i=1\wedge b_i=1)\\
1-p, & \textrm{if $b_1=b_2=\ldots=b_m=0$} \\
0, & \textrm{otherwise}\end{array}\right.$ \\[7mm]
$\mathsf{VIRTUALIZED}$ & 
$p_{b_1,b_2,\ldots, b_m}=\left\{ 
\begin{array}{ll} 
p \prod_{i=1}^m f(b_i, n_i), 
& 
\textrm{if } \exists i\in\{1,2,\ldots,m\}.b_i\neq 0\\[2mm]
p\prod_{i=1}^m f(b_i, n_i) + (1-p), & \textrm{if $b_1=b_2=\ldots=b_m=0$}
\end{array}
\right.$ \\[7mm] 
$\mathsf{VIRTUALIZED\textrm{-}M}$ & 
$p_{b_1,b_2,\ldots, b_m}=\left\{ 
\begin{array}{ll} 
p \prod_{i=1}^m f(b_i, n_i) +(1-p)p_\mathsf{detect} \prod_{i=1}^m g(b_i, n_i), 
& 
\textrm{if } \exists i\in\{1,2,\ldots,m\}.b_i\neq 0\\[2mm]
p \prod_{i=1}^m f(b_i, n_i) + (1-p)p_\mathsf{detect}\prod_{i=1}^m g(b_i, n_i) + (1-p)(1-p_\mathsf{detect}), & \textrm{if $b_1=b_2=\ldots=b_m=0$}
\end{array}
\right.$ \\[3mm]
\midrule
\\[-2mm]
\multicolumn{2}{l}{$\textrm{with }
f(b_i,n_i)\!=\!\left\{\!\!\!
\begin{array}{ll}
(1-p_\mathsf{VM})^{n_i}, & \textrm{if } b_i=0\\[1mm]
n_ip_\mathsf{VM}(1-p_\mathsf{VM})^{n_i-1}, & \textrm{if } b_i=1\\[1mm]
1-f(0,n_i)-f(1,n_i), & \textrm{if $b_i=2+$}\\
\end{array}\!\!
\right.$; 
$
\;g(b_i,n_i)\!=\!\left\{\!\!\!
\begin{array}{ll}
\left(
\frac{(1-p_\mathsf{migrate})(1-r)+p_\mathsf{migrate}(1-p_\mathsf{VM})}{1-(1-p_\mathsf{migrate})r}
\right)^{n_i}, & \textrm{if } b_i=0\\[2mm]
n_i\frac{p_\mathsf{migrate}p_\mathsf{VM}}{1-(1-p_\mathsf{migrate})r}\left(\frac{(1-p_\mathsf{migrate})(1-r)+p_\mathsf{migrate}(1-p_\mathsf{VM})}{1-(1-p_\mathsf{migrate})r}\right)^{n_i-1}, & \textrm{if } b_i=1\\[2mm]
1-g(0,n_i)-g(1,n_i), & \textrm{if $b_i=2+$}\\
\end{array}
\right.\!\!\!$
}\\
\bottomrule
\end{tabular}

\vspace*{-1mm}
\end{table*}

\begin{figure}
\centering
\includegraphics[width=\hsize]{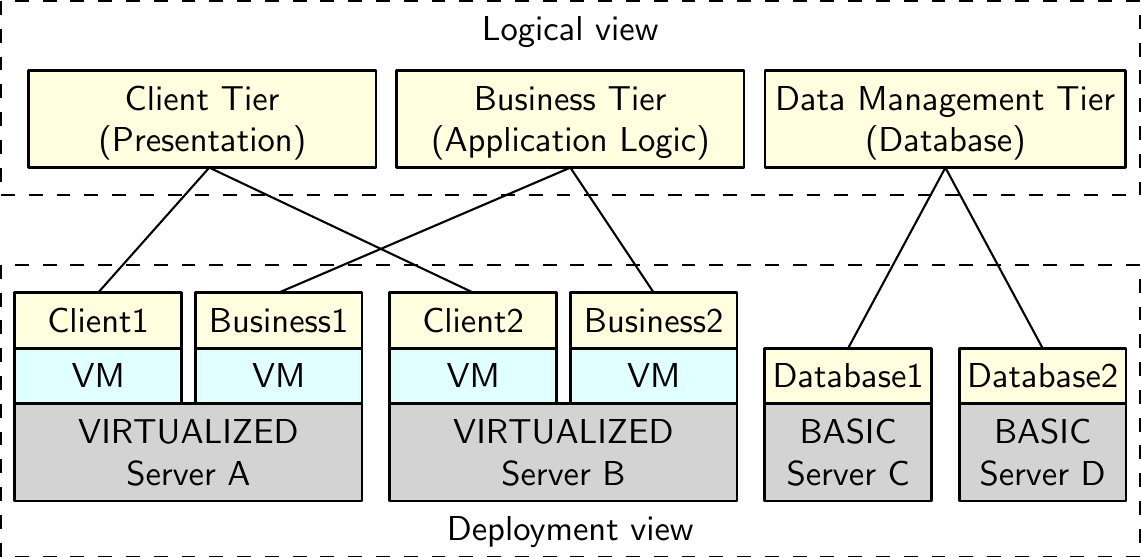}

\vspace*{-1.5mm}
\caption{Three-tier system deployed across four servers \label{fig:three-tier}}

\vspace*{-3mm}
\end{figure}

\begin{theorem}
The $p_{b_1,b_2,\ldots,b_m}$ expressions from Table~\ref{tab:server_expressions} specify the probabilities that $b_1,b_2,\ldots,b_m$ instances of tiers $1,2,\ldots,m$ remain operational on a $\mathsf{BASIC}$, $\mathsf{VIRTUALIZED}$ and $\mathsf{VIRTUALIZED}$-$\mathsf{M}$ server, respectively.
\end{theorem}

\begin{example}
\label{ex:multitier}
We consider a three-tier system adapted from \cite{calinescu2012compositional,DBLP:conf/cbse/JohnsonCK13}, and comprising client, business and data management tiers. We assume that there are two instances of each tier, and that these instances are deployed on four servers as shown in Fig.~\ref{fig:three-tier}. To compute systems of closed-form expressions for the reliability properties $P_\mathsf{FAIL}$ and $P_\mathsf{SPF}$ introduced earlier in this section, we built the pattern-annotated Markov chain from Fig.~\ref{fig:three-tier-MC}. 
The states of this MC correspond to different numbers of tier instances being  operational at different stages of the analysis: a state labelled `$x,y,z$' models the scenario where the system has $x$ client-tier instances, $y$ business-tier instances and $z$ database-tier instances active. In the initial state, $x=y=z=2$, and the values of the three variables may decrease as the potential failures of each of servers A--D are modelled in four successive stages. The MC states within these stages are annotated with the relevant server modelling patterns, i.e., \textsf{VIRTUALIZED}$(1,1,p^A,p_\mathsf{VM}^A)$ for server A, \textsf{VIRTUALIZED}$(1,1,p^B,p_\mathsf{VM}^B)$ for server B, \textsf{BASIC}$(1,p^C)$ for server C, and \textsf{BASIC}$(1,p^D)$ for server D. To keep the model simple, all final states with zero instances within at least one tier are joined together into a single state (labelled `FAIL'); and all final non-FAIL states with only one instance within at least one tier are combined into a ``single point of failure'' (`SPF') state. The model includes no state transitions with probabilities $p^A_{b_1,b_2}$, $p^B_{b_1,b_2}$, $p^C_{b_1}$ and $p^D_{b_1}$ with $b_1=2+$ or $b_2=2+$ because all these probabilities are zero: the servers cannot end up running two or more instances of the same tier as at most one instance of each tier is deployed on each server to start with.

\begin{figure}
\centering
\includegraphics[width=\hsize]{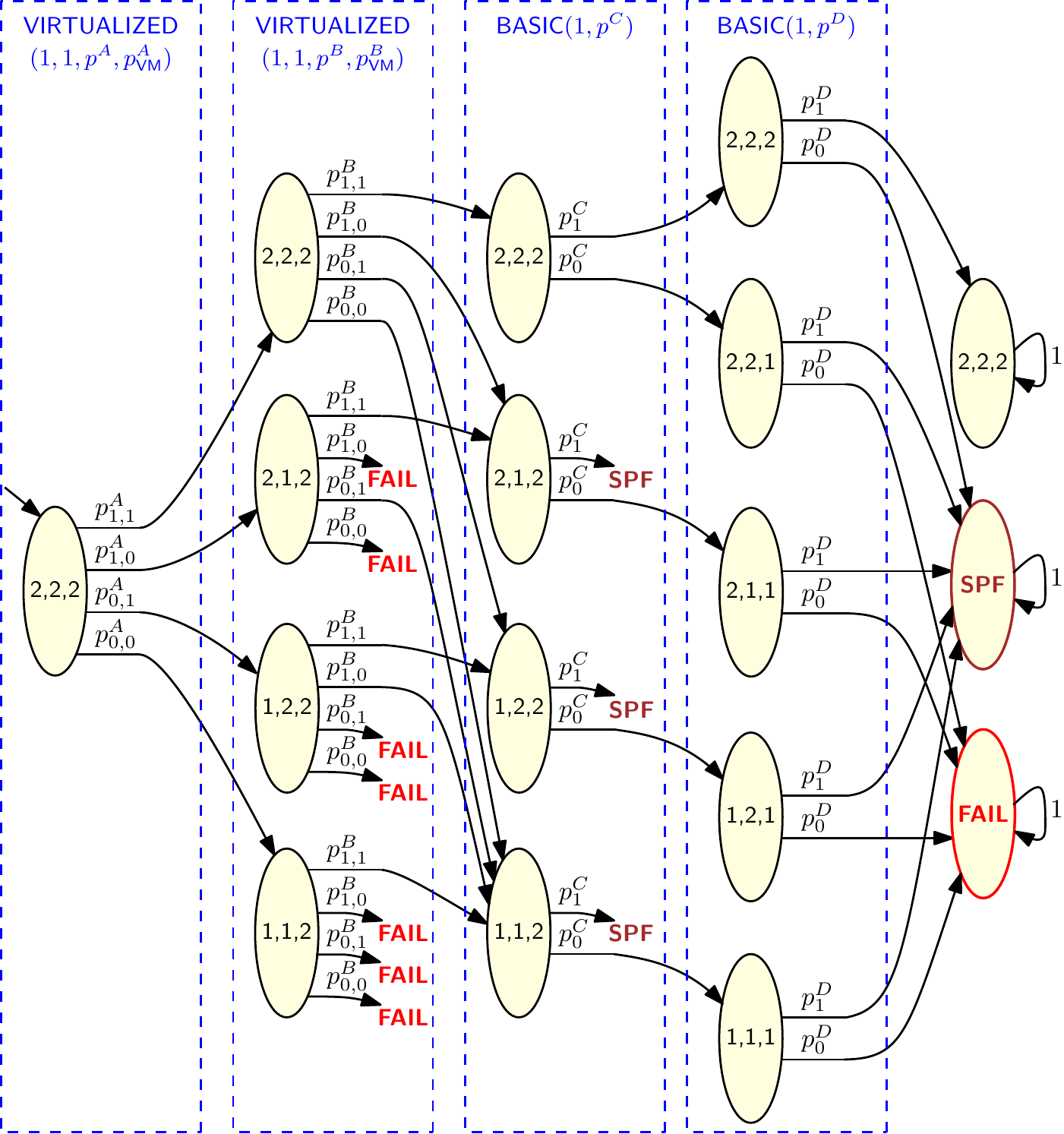}
\caption{Pattern-annotated Markov chain for the three-tiered system deployment  
from Fig.~\ref{fig:three-tier} \label{fig:three-tier-MC}}

\vspace*{-7mm}
\end{figure}

Given this pattern-annotated parametric MC and the repository from Table~\ref{tab:server_expressions}, ePMC computes the set of formulae from Table~\ref{table:multi-tier-example} for the properties $P_\mathsf{FAIL}=\mathcal{P}_{=?}[\mathsf{F}\; \mathsf{FAIL}]$ and $P_\mathsf{SPF}=\mathcal{P}_{=?}[\mathsf{F}\; \mathsf{SPF}]$. The first two formulae from this table were obtained using Storm \cite{Dehnert2017} to verify the parametric MC from Fig.~\ref{fig:three-tier-MC}, and the other formulae were obtained from the repository in Table~\ref{tab:server_expressions}.

\begin{table}
\caption{Probability of failure ($P_\mathsf{FAIL}$) and probability of single point of failure ($P_\mathsf{SPF}$) for the three-tier system from Fig.~\ref{fig:three-tier} \label{table:multi-tier-example}}
\begin{tabular}{p{8.45cm}} 
\toprule
$
\!\!\!\!\begin{array}{l}
\!\!\!P_\mathsf{FAIL}
=p_{0,0}^Ap_{1,1}^Bp_0^Cp_0^D+p_{0,0}^A+p_{0,1}^Ap_{1,0}^Bp_0^Cp_0^D+p_{1,1}^Ap_{0,0}^Bp_0^Cp_0^D+\\
\;p_{0,1}^Ap_{1,1}^Bp_0^Cp_0^D+p_{1,1}^Ap_{1,1}^Bp_0^Cp_0^D+p_{1,1}^Ap_{1,0}^Bp_0^Cp_0^D+p_{0,1}^Ap_{0,1}^B+\\
\;p_{1,0}^Ap_{1,1}^Bp_0^Cp_0^D+p_{1,0}^Ap_{1,0}^B+p_{1,0}^Ap_{0,0}^B+p_{0,1}^Ap_{0,0}B+p_{1,1}^Ap_{0,1}^Bp_0^Cp_0^D+\\
\;p_{1,0}^Ap_{0,1}^Bp_0^Cp_0^D-p_{0,0}^Ap_{1,1}^B\\[2mm]
\!\!\!P_\mathsf{SPF}=
p_{0,0}^Ap_{1,1}^Bp_0^Cp_1^D\!+p_{0,1}^Ap_{1,0}^Bp_0^Cp_1^D\!+p_{1,0}^Ap_{0,1}^Bp_1^C\!+p_{1,0}^Ap_{0,1}^Bp_0^Cp_1^D\!+\\
\;p_{0,1}^Ap_{1,0}^Bp1C+p_{1,1}^Ap_{0,0}^Bp1C+p_{0,1}^Ap_{1,1}^Bp_0^Cp_1^D+p_{1,1}^Ap_{0,1}^Bp_1^C+\\
\;p_{1,1}^Ap_{1,1}^Bp_1^Cp_0^D+p_{1,1}^Ap_{1,1}^Bp_0^Cp_1^D+p_{1,1}^Ap_{1,0}^Bp1C+p_{1,1}^Ap_{1,0}^Bp_0^Cp_1^D+\\
\;p_{1,0}^Ap_{1,1}^Bp_0^Cp_1^D+p_{0,1}^Ap_{1,1}^Bp_1^C+p_{1,0}^Ap_{1,1}^Bp1C+p_{1,1}^Ap_{0,1}^Bp_0^Cp_1^D+\\
\;p_{1,1}^Ap_{0,0}^Bp_0^Cp_1^D+p_{0,0}^Ap_{1,1}^Bp_1^C\\[2mm]
\!\!\!p^A_{1,1}=p^A(p^A_\mathsf{VM})^2\qquad\qquad\quad\,  p^A_{1,0}=p^Ap^A_\mathsf{VM}(1-p^A_\mathsf{VM})\\[2mm]
\!\!\!p^A_{0,1}=p^Ap^A_\mathsf{VM}(1-p^A_\mathsf{VM})\qquad  p^A_{0,0}=(1-p^A)+p^A(1-p^A_\mathsf{VM})^2\\[2mm]
\!\!\!p^B_{1,1}=p^B(p^B_\mathsf{VM})^2\qquad\qquad\quad\,  p^B_{1,0}=p^Bp^B_\mathsf{VM}(1-p^B_\mathsf{VM})\\[2mm]
\!\!\!p^B_{0,1}=p^Bp^B_\mathsf{VM}(1-p^B_\mathsf{VM})\qquad  p^B_{0,0}=(1-p^B)+p^B(1-p^B_\mathsf{VM})^2\\[2mm]
\!\!\!p^C_1=p^C\quad\;\,  p^C_0=1-p^C\quad\;\, p^D_1=p^D\quad\;\, p^D_0=1-p^D
\end{array}
$\\
\bottomrule
\end{tabular}

\vspace*{-2mm}
\end{table}
\end{example}

\section{Evaluation \label{sect:evaluation}}

We carried out extensive experiments to compare the feasibility, scalability and efficiency of ePMC to those of the model checkers PRISM, Storm and PARAM. All experiments were performed on a Ubuntu-16 server with i7-4770@3.40GHz $\times$ 8 processors and 16GB of memory, on which we installed the latest versions of the three model checkers downloaded from their websites and our ePMC pattern-aware parametric model checker from Section~\ref{sect:tool}. To ensure the reproducibility of our results, we made the models and verified properties from our experiments available at  \url{https://www.cs.york.ac.uk/tasp/ePMC/}.

To also assess the generality of our method, we evaluated it for both the service-based systems domain introduced in Section~\ref{sect:sbs} and the multi-tier architectures domain presented in Section~\ref{sect:multitier}. The experimental results for these two domains are reported in the next two sections.

\subsection{Service-based systems domain \label{subsect:eval-sbs}}

\begin{figure}
\centering
\includegraphics[width=0.8\hsize]{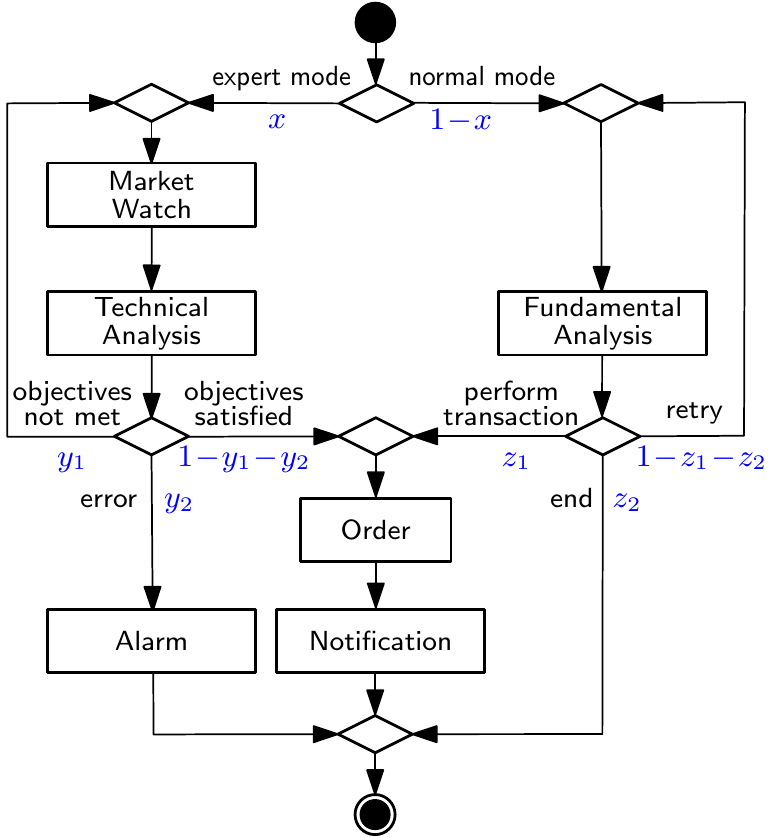}

\vspace*{-2mm}
\caption{Foreign exchange system from the SBS domain\label{fig:fx}}

\vspace*{-3mm}
\end{figure}

We evaluated ePMC by using it to analyse  a six-component service-based system initially introduced in \cite{DBLP:conf/kbse/GerasimouTC15} and also used in \cite{Calinescu2017,DBLP:journals/ase/GerasimouCT18}. This system implements a workflow used to carry out foreign exchange (FX) trading transactions as illustrated by the UML activity diagram in Fig.~\ref{fig:fx}. Traders can use the FX system in ``normal'' or ``expert'' operation modes. In its normal mode, the system uses a Fundamental Analysis component to decide whether a transaction should be performed, or the fundamental analysis should be retried, or the normal-mode session should be ended. Performing a transaction involves using an Order component to carry out the operation, and is followed by the invocation of a Notification component to inform the trader about the outcome of the transaction.
In its expert mode, the system uses a Market Watch component to obtain exchange market data that is then processed by a Technical Analysis component. The result of this analysis may satisfy a set of trader-specified objectives (in which case a transaction is performed), may not meet these objectives (so the Market Watch is reinvoked for an update) or may be erroneous (in which case an Alarm component is used to warn the trader). In our experiments, we assumed that the probabilities that annotate the decision points from the diagram in Fig.~\ref{fig:fx} (i.e., the operational profile of the FX system) were unknown parameters $x$, $y_1$, $y_2$, $z_1$ and $z_2$.

To evaluate ePMC, we considered multiple ways in which the six FX components could be implemented using the SBS modelling patterns from Table~\ref{tab:sbs_patterns} with between one and five functionally-equivalent services per component. To analyse this large set of alternative designs using ePMC, we developed a pattern-annotated parametric MC similar to the MC from Fig.~\ref{fig:running-pmc}c(ii) but modelling the FX system. To analyse the same designs with the model checkers PRISM, Storm and PARAM, we obtained an individual ``monolithic'' model for each design by using a dedicated parametric MC generator that we implemented for this purpose. 

Three QoS properties of the system were analyzed: (P1)~the probability of successful completion; (P2)~the expected execution time; and (P3)~the expected cost. Table~\ref{tab:pmc_time} compares the PMC time required to produce the sets of formulae for these three properties using ePMC to the time required to produce a single formula for each property using PRISM, Storm and PARAM. With the exception of the last row, the results correspond to experiments in which every FX component used the same pattern ($\mathsf{SEQ}$, $\mathsf{PAR}$, etc.) and the same number of services. 

\begin{table*}
	\caption{Parametric model checking time (seconds) for the FX service-based system}
	\label{tab:pmc_time}
	
	\vspace*{-2mm}
	\centering
	\def\arraystretch{0.7}
	\begin{tabular}{cccccccccccccc} 
		\toprule
		& &  \multicolumn{3}{c}{\textbf{ePMC}} &  \multicolumn{3}{c}{\textbf{PRISM}} &  \multicolumn{3}{c}{\textbf{Storm}} &  \multicolumn{3}{c}{\textbf{PARAM}} \\
		\textbf{Pattern} & \textbf{\#services} & P1 & P2 & P3 & P1 & P2 & P3 & P1 & P2 & P3 & P1 & P2 & P3\\
		\midrule 
		$\mathsf{SEQ}$ & 1 & 0.33  & 0.34 & 0.33 & 0.32 & 8.10 & 8.51 & 0.06 & 0.077 & 0.075 & 0.012 & 0.862 & 0.883 \\ \midrule
		$\mathsf{SEQ}$ & 2   & 0.35 & 0.35 & 0.38 & 32.87 & M  & M   & 3.61 & 7.61 & 7.59 & 5.24 & T & T \\
		$\mathsf{SEQ}$ & 3 &  0.34 & 0.34 & 0.34 & M   &    -   &    -  & T & T & T & T &  - &  -  \\ 
		$\mathsf{SEQ}$ & 4 &  0.36 & 0.35 & 0.36 &   -    &  -     &   -    & - & -  &  - &   - &   -   &  -\\ 
		$\mathsf{SEQ}$ & 5 &  0.34 & 0.34 & 0.34 &   -    &  -     &   -    & - & -  &  - &  - &   -   &  -\\ \midrule
		
		$\mathsf{PAR}$ & 2 & 0.33 & 0.34 & 0.34 & 8.75 & M   & M   & 1.69 & 8.54 & 3.92 & 5.14 & T & T \\ 
		$\mathsf{PAR}$ & 3 & 0.33 & 0.36 & 0.34 & M   &   -    &    -   & T   & T & T & T & - & -  \\ 
		$\mathsf{PAR}$ & 4 & 0.48 & 0.35 & 0.35 &   -    &    -   &   -    &  -  &  -  & - & - & - & - \\ 
		$\mathsf{PAR}$ & 5 & 0.34 & 0.35 & 0.35 &   -    &  -     &   -    & - & -  &  - &  - &   -   &  -\\ \midrule
		
		$\mathsf{PROB}$ & 2 & 0.34 & 0.34 & 0.35 & M   & M   & M   & 0.46 & 0.89 & 0.87 & T & T & T \\ 
		$\mathsf{PROB}$ & 3 & 0.35 & 0.35 & 0.34 &    -   &    -   &  -     & 2.81 & 4.34 & 4.30 &  -    &   -   &  - \\ 
		$\mathsf{PROB}$ & 4 & 0.36 & 0.35 & 0.35 &   -    &     -  &   -    & 48.23  & 50.08  & 50.33  &      - &   -   & - \\ 
		$\mathsf{PROB}$ & 5 & 0.35 & 0.37 & 0.35 &   -    &  -     &   -    & 545.33 &  395.12  &  387.27  &   -  &   -   &  -\\ \midrule
		
		$\mathsf{SEQ\_R}$ & 2 & 0.35 & 0.36 & 0.35& M   & M   & M   & 830.24  & T  & T  & T & T & T \\
		$\mathsf{SEQ\_R}$ & 3 & 0.37 & 0.36 & 0.36 &   -    &    -   &    -   &    -  &    -   &    -   &  -    &   -   &  -\\ 
		$\mathsf{SEQ\_R}$ & 4 & 0.36 & 0.36 & 0.37 &   -    &    -   &    -   &    -  &    -   &    -   &  -    &   -   &  -\\ 
		$\mathsf{SEQ\_R}$ & 5 & 0.36 & 0.37 & 0.35 &   -    &    -   &    -   &    -  &    -   &    -   &  -    &   -   &  -\\ \midrule
		
		$\mathsf{SEQ\_R1}$ & 2 &  0.36 & 0.35 & 0.37 & M   & M   & M   & T  & T  & T  & T & T & T \\
		$\mathsf{SEQ\_R1}$ & 3 &  0.35 & 0.35 & 0.37 &   -    &    -   &    -   &    -  &    -   &    -   &  -    &   -   &  -\\ 
		$\mathsf{SEQ\_R1}$ & 4 &  0.35 & 0.36 & 0.37 &    -   &     -  &    -   &   -    &    -   &    -   &   -   &   -   &  -\\ 
		$\mathsf{SEQ\_R1}$ & 5 &  0.36 & 0.39 & 0.36 &   -    &  -     &   -    & - & -  &  - &  - &   -   &  -\\ \midrule

		$\mathsf{PAR\_R}$ & 2 & 0.35 & 0.36 & 0.36& M   & M   & M   & 566.18  & T  & T  & T & T & T\\
		$\mathsf{PAR\_R}$ & 3 & 0.36 & 0.36 & 0.36 &   -    &    -   &    -   &  -  &     -  &   -    & -&- &-\\ 
		$\mathsf{PAR\_R}$ & 4 & 0.35 & 0.37 & 0.38 &   -    &    -   &    -   &  -  &     -  &   -    & -&- &-\\ 
		$\mathsf{PAR\_R}$  & 5 & 0.34 & 0.37 & 0.35 &   -    &    -   &    -   &  -  &     -  &   -    & -&- &-\\ \midrule
		
		$\mathsf{PROB\_R}$ & 2& 0.37 & 0.35 & 0.38 & M   & M   & M   & 70.27 & 68.90  & 69.42  & T & T & T \\
		$\mathsf{PROB\_R}$ & 3 & 0.37 & 0.36 & 0.35 &   -    &    -   &    -   & T &   T    &     T  &- &- & - \\
		$\mathsf{PROB\_R}$ & 4 & 0.35 & 0.36 & 0.37 &   -    &    -   &    -   &  -  &     -  &   -    & -&- &- \\
		$\mathsf{PROB\_R}$ & 5 & 0.37 & 0.37 & 0.36 &   -    &  -     &   -    & - & -  &  - &  - &   -   &  -\\
		\midrule
		
		$\mathsf{PROB\_R1}$  & 2 & 0.36 & 0.37 & 0.36& M   & M   & M   & T  & T  & T  & T & T & T\\
		$\mathsf{PROB\_R1}$ & 3 & 0.35 & 0.36 & 0.37 &   -    &    -   &    -   &  -  &     -  &   -    & -&- &-\\ 
		$\mathsf{PROB\_R1}$ & 4 & 0.35 & 0.36 & 0.35 &   -    &    -   &    -   &  -  &     -  &   -    & -&- &-\\ 
		$\mathsf{PROB\_R1}$ & 5 & 0.36 & 0.36 & 0.36 &   -    &    -   &    -   &  -  &     -  &   -    & -&- &-\\ \midrule
		
		20~random & min & 0.31 & 0.31 & 0.31& T & T & T & 2.25 & 2.79 & 1.12 & T & T & T \\
		combinations & max & 0.54 & 0.34 & 0.37& & & & 520.45 & 817.58 & 661.80 \\
		of 2/3-service & mean & 0.33 & 0.32 & 0.32 & & & & 52.12 &  88.41 & 67.18 \\
		\textsf{SEQ/PAR/PROB} & stdev & 0.06 & 0.01 & 0.01 & & & & 112.93 & 178.01 & 148.62 \\
		
		\bottomrule
		\\[-1mm]
		\multicolumn{12}{l}{M=out of memory, T=timeout (no result returned within 15 minutes), --=experiment skipped as PMC of smaller model failed}\\
	\end{tabular}
\end{table*}

\begin{figure*}
\centering
\includegraphics[width=0.32\linewidth]{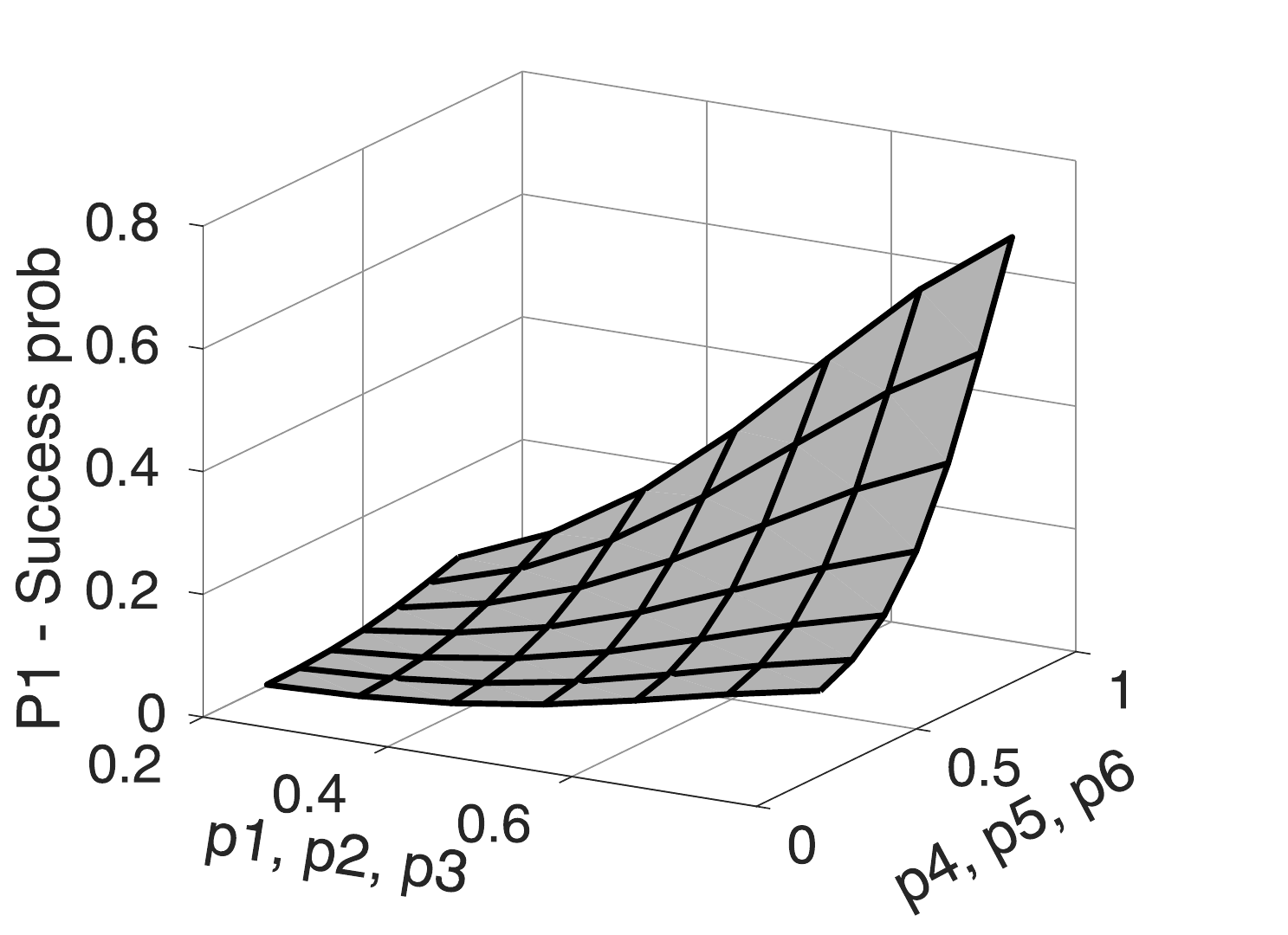}
\includegraphics[width=0.32\linewidth]{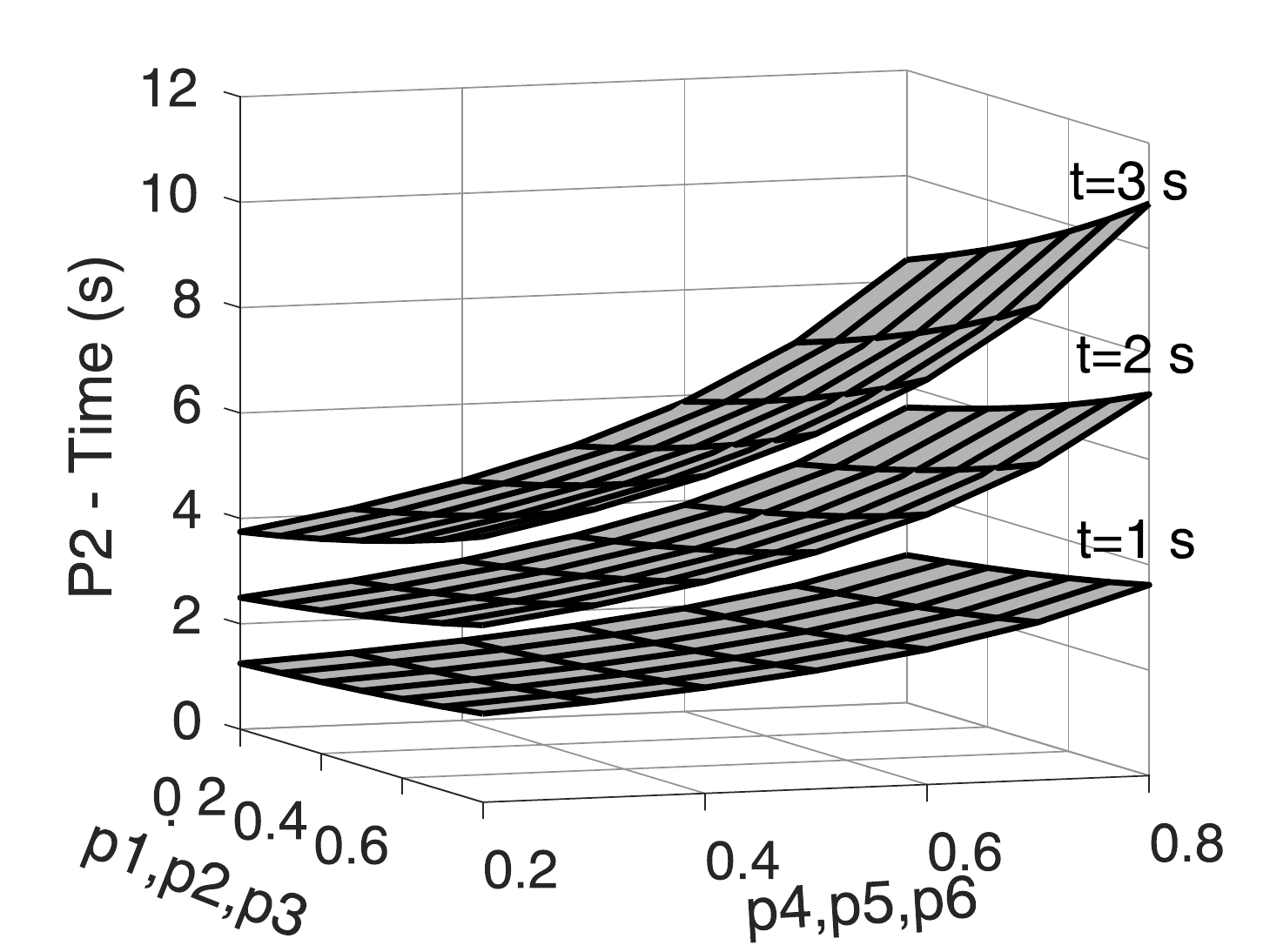}\hspace*{4mm}
\includegraphics[width=0.32\linewidth]{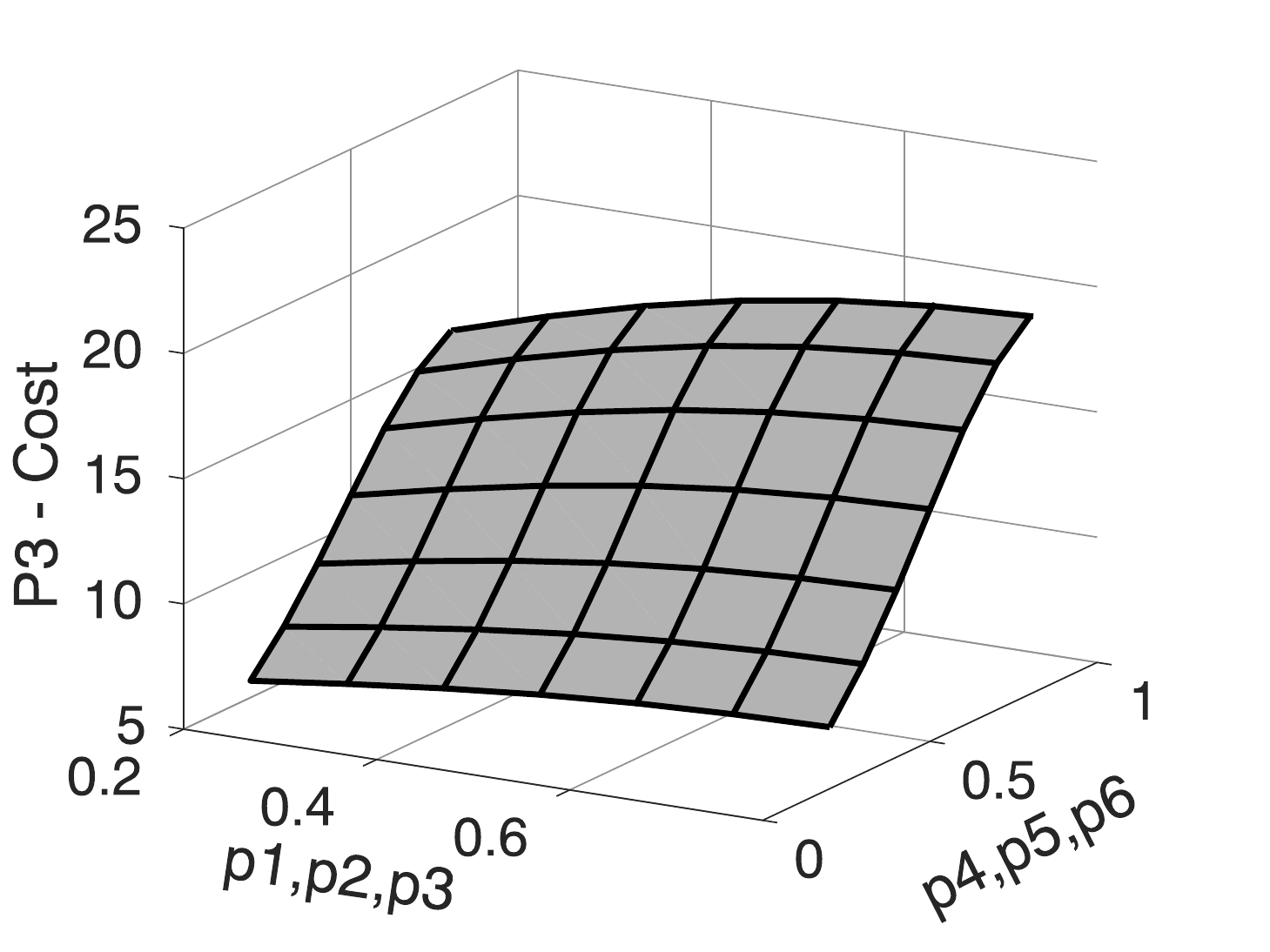}

\vspace*{-1.5mm}
\caption{QoS analysis showing increase in FX success probability and (as fewer FX sessions fail halfway) in expected execution time and cost when the reliability $p_i$ of (all) services used for the $i$-th FX component, $1\leq i\leq 6$, increases; for simplicity, all services are assumed to have the same mean execution time (1s, 2s or 3s) and the same cost of 1 
\label{fig:fx-plots} }

\vspace*{-2mm}
\end{figure*}

Table~\ref{tab:pmc_time} shows that the PMC time required to analyse the three properties using ePMC is always better, and typically orders of magnitude smaller, than the PMC times of PRISM, Storm and PARAM (except for the trivial case when a single service is used for each SBS component, cf.~row~1). Moreover, the three tools ran out of memory or timed out when components used four (and sometimes even two or three) services, with the exception of the Storm analyses of the PROB pattern, which were all completed. The reason for this is that PROB is by far the SBS modelling pattern with the simplest QoS-property expressions, i.e., just linear combinations of the service parameters, as shown in Table~\ref{tab:sbs_expressions}. 
Note also that ePMC analysis times are almost identical irrespective of the property analysed and of the pattern and number of services used. This is because the time required to run our ePMC tool is dominated by the time used to read the files containing the ePMC repository of QoS-property expressions and the annotated MC, to start Storm and to parse the MC, all of which do not depend on the analysed property or the patterns from the model annotations.

The last row from Table~\ref{tab:pmc_time} reports the minimum, maximum and mean PMC time and the standard deviation over 20 experiments in which the pattern and number of services (two or three) used for each component were chosen randomly and independently of those of the other components. We only used the patterns SEQ, PAR and PROB in these experiments so that at least Storm could complete the analysis, albeit with PMC times much longer than ePMC; PRISM and PARAM timed out in all 20~experiments.

To assess the efficiency of evaluating ePMC-generated expressions, we plotted graphs of the three QoS properties of the FX system using both the sets of formulae generated by our ePMC tool and the ``monolithic'' formulae generated by Storm (the best performing of the current model checkers in our experiments). Fig.~\ref{fig:fx-plots} shows three such graphs, generated with Matlab. These graphs correspond to the following fixed values for the FX operational profile parameters (which can be obtained in practice from system logs): $x=0.66$, $y_1=0.61$, $y_2=0.11$, $z_1=0.27$ and $z_2=0.53$. Table~\ref{table:eval-comparison} shows that ePMC yields far smaller and more efficient to evaluate formulae than traditional PMC. The only system instance for which the Storm formula size and graph generation time are comparable to (but still larger than) those of ePMC corresponds to PROB, i.e., the simplest SBS modelling pattern as discussed earlier in this section. For the other patterns from Table~\ref{table:eval-comparison}, the ePMC sets of formulae are several orders of magnitude smaller and faster to evaluate than the Storm formulae.

\begin{table}
\centering
\caption{Comparison of ePMC and Storm formulae sizes and graph generation times for the graphs from Fig.~\ref{fig:fx-plots} and system instances using the same SBS modelling pattern (with three services) for each FX component \label{table:eval-comparison}}

\vspace*{-1mm}
\def\tabcolsep{5pt}
	\begin{tabular}{cccccc} 
		\toprule
		& \textbf{SBS modelling\hspace*{-3mm}} & \multicolumn{2}{c}{\textbf{ePMC formulae}} & \multicolumn{2}{c}{\textbf{Storm formula}}\\
		\textbf{\hspace*{-2mm}Graph\hspace*{-3mm}} & \textbf{pattern} & \textbf{\#operations\hspace*{-1mm}} & \textbf{time} & \textbf{\#operations\hspace*{-1mm}} & \textbf{time} \\
		\midrule
		P1 & PROB\_R   & 287 & 2.6s & 285425 & 8.2 hours\\
		P2 & PROB   & 174 & 2.9s & 9686 & 16.9s \\
		P3 & PAR   & 198 & 1.3s & 171166 & 2.5 hours \\
		\bottomrule
	\end{tabular}

\vspace*{-5mm}
\end{table}

\begin{table*}
\caption{Parametric model checking time (seconds or T=15-minute timeout) for eight deployments of a three-tier system \label{table:multi-tier-deployments}}

\vspace*{-2mm}
\centering
\def\arraystretch{1}
\def\tabcolsep{4pt}
	\begin{tabular}{lllllllllcccccccc} 
		\toprule
		& \multicolumn{4}{c}{\textbf{Tier instances$^\dagger$}}$\!\!\!$ & \multicolumn{4}{c}{\textbf{Server type$^\ddagger$ $\mid$ Instances of tiers 1, 2, 3}} & \multicolumn{2}{c}{\textbf{ePMC}}  & \multicolumn{2}{c}{\textbf{PRISM}}  & \multicolumn{2}{c}{\textbf{Storm}} & \multicolumn{2}{c}{\textbf{PARAM}} \\[1mm]
		\textbf{ID}  & \textbf{T1$\;$} & \textbf{T2$\;$} & \textbf{T3$\;$} & \textbf{Total} & \textbf{Server A} & \textbf{Server B} & \textbf{Server C} & \textbf{Server D} & $P_\mathsf{FAIL}$ & $P_\mathsf{SPF}$ & $P_\mathsf{FAIL}$ & $P_\mathsf{SPF}$ & $P_\mathsf{FAIL}$ & $P_\mathsf{SPF}$ & $P_\mathsf{FAIL}$ & $P_\mathsf{SPF}$ \\
		\midrule
		D1$^\textsf{\#}$ & 
		2 & 2 & 2 & 6 & \textsf{V \hspace*{3.1mm}$\mid$ 1,1,0} & \textsf{V \hspace*{3.1mm}$\mid$ 1,1,0} & \textsf{B \hspace*{3.1mm}$\mid$ 0,0,1} & \textsf{B \hspace*{3.1mm}$\mid$ 0,0,1} &0.26&0.26&0.39&0.15&0.55&0.10&0.06&0.03 \\
		D2 & 2 & 2 & 2 & 6 & \textsf{V-M $\mid$ 1,1,0} & \textsf{V-M $\mid$ 1,1,0} & \textsf{B \hspace*{3.1mm}$\mid$ 0,0,1} & \textsf{B \hspace*{3.1mm}$\mid$ 0,0,1}&0.25&0.26&3.14&7.33&5.08&4.92&12.88&33.06 \\
		D3 & 4 & 4 & 2 & 10 & \textsf{V \hspace*{3.1mm}$\mid$ 2,1,0} & \textsf{V \hspace*{3.1mm}$\mid$ 2,1,0} & \textsf{V $\hspace*{3.1mm}\mid$ 0,1,1} & \textsf{V \hspace*{3.1mm}$\mid$ 0,1,1}&0.26&0.26&2.50&1.86&1.00&1.01&0.53&0.53 \\
		D4 & 4 & 4 & 2 & 10& \textsf{V-M $\mid$ 2,1,0} & \textsf{V-M $\mid$ 2,1,0} & \textsf{V-M $\mid$ 0,1,1} & \textsf{V-M $\mid$ 0,1,1}&0.27&0.27&T&T&T&T&T&T \\
		D5 & 8 & 8 & 4 & 20 & \textsf{V \hspace*{3.1mm}$\mid$ 4,2,0} & \textsf{V \hspace*{3.1mm}$\mid$ 4,2,0} & \textsf{V \hspace*{3.1mm}$\mid$ 0,2,2} & \textsf{V \hspace*{3.1mm}$\mid$ 0,2,2}  &0.29 &0.31&742.85 &1074.59&25.19 &26.92&12.78 &14.11\\
		D6 & 8 & 8 & 4 & 20 & \textsf{V-M $\mid$ 4,2,0} & \textsf{V-M $\mid$ 4,2,0} & \textsf{V-M $\mid$ 0,2,2} & \textsf{V-M $\mid$ 0,2,2} &0.29 &0.30 &T&T&T&T&T&T  \\
		D7 & 16 & 16 & 8 & 40 & \textsf{V \hspace*{3.1mm}$\mid$ 8,4,0} & \textsf{V \hspace*{3.1mm}$\mid$ 8,4,0} & \textsf{V \hspace*{3.1mm}$\mid$ 0,4,4} & \textsf{V \hspace*{3.1mm}$\mid$ 0,4,4}  &0.29 &0.29&T &T&T &T&T &T\\
		D8 & 16 & 16 & 8 & 40 & \textsf{V-M $\mid$ 8,4,0} & \textsf{V-M $\mid$ 8,4,0} & \textsf{V-M $\mid$ 0,4,4} & \textsf{V-M $\mid$ 0,4,4} &0.30 &0.30 &T&T&T&T&T&T \\
		\bottomrule
		\multicolumn{16}{l}{$^\dagger$ T1=Tier 1 instances; T2=Tier 2 instances; T3=Tier 3 instances}\\
		\multicolumn{16}{l}{$^\ddagger$ B=BASIC; V=VIRTUALIZED; V-M=VIRTUALIZED-M}\\
		\multicolumn{16}{l}{$^\textsf{\#}$ Deployment used in Example~\ref{ex:multitier}}
	\end{tabular}

\vspace*{-2mm}
\end{table*}

\subsection{Multi-tier software architectures domain \label{subsect:eval-multi}}

We evaluated ePMC within this domain by using it to analyse the properties $P_\mathsf{FAIL}$ and $P_\mathsf{SPF}$ from Section~\ref{sect:multitier} for  eight four-server deployments of a three-tier system. The characteristics of these deployments and the time taken by the parametric model checking of the two properties are shown in Table~\ref{table:multi-tier-deployments}. 

As for the SBS domain, the ePMC model checking time is largely unaffected by the system size, remaining under~1s when the total number of tier instances increases from six instances for deployments D1 and D2 to 40 instances for deployments D7 and D8. In contrast, the model checking time for PRISM, Storm and PARAM increases rapidly with the system size. As D1, D3, D5 and D7 use only the simpler, loop-free deployment patterns \textsf{BASIC} and \textsf{VIRTUALIZED}, the three model checkers can successfully analyse deployments D1, D3 and D5. 
However, the analysis times of these tools are already orders of magnitude larger than those of ePMC for the larger deployment D5 (and their analyses of deployment D7 time out). The better efficiency of ePMC is even clearer for deployments D2, D4, D6 and D8, which use the more complex deployment pattern \textsf{VIRTUALIZED-M}---out of these  deployments, only D2 can be successfully analysed by PRISM, Storm and PARAM.

\begin{table}
	\caption{Combined size of $P_\mathsf{FAIL}$ and $P_\mathsf{SPF}$ formulae (\#operations or T=timeout) for the parametric model checking experiments from Table~\ref{table:multi-tier-deployments} \label{table-multi-tier-formulae-size}}
	
	\vspace*{-2mm}
	\centering
	\begin{tabular}{ccccc} \toprule
		\textbf{ID} & \textbf{ePMC} & \textbf{PRISM} & \textbf{Storm} & \textbf{PARAM}\\ 
		\midrule
		D1&  143 & 204 & 258& 240\\
		D2&  189 & 30584 & 34719 & 33667\\
		D3& 1688 & 1892 & 2234 & 2124\\
		D4& 1868 & T&T & T\\	
		D5&  9082&  21952& 25394& 24248\\	
		D6&  9404& T& T&T\\	
		D7&  9086& T& T&T\\	
		D8&  9412& T&T &T\\	
		\bottomrule
	\end{tabular}

	\vspace*{-2mm}
\end{table}

Finally, Table~\ref{table-multi-tier-formulae-size} shows the combined sizes of the $P_\mathsf{FAIL}$ and $P_\mathsf{SPF}$ formulae generated by ePMC and by the current model checkers for the deployments from Table~\ref{table:multi-tier-deployments}. As for the SBS domain, the ePMC formulae are always smaller than those produced by the current model checkers. Moreover, they are over two orders of magnitude smaller for deployment D2, which is the only deployment that uses the more complex modelling pattern \textsf{VIRTUALIZED-M} and that 
PRISM, Storm and PARAM can analyse.

\section{Related work \label{sect:related}}

Since its introduction in Daws' seminal work \cite{Daws:2004:SPM:2102873.2102899} in 2004, parametric model checking has underpinned the development of a vast array of methods for the modelling and analysis of software and other computer-based systems. These include methods for comparing alternative system designs \cite{DBLP:conf/splc/GhezziS11,DBLP:journals/infsof/GhezziS13}, sensitivity analysis \cite{DBLP:journals/tse/FilieriTG16}, parameter synthesis \cite{calinescu2017synthesis,Dehnert2015,hahn2011synthesis,DBLP:journals/jss/CalinescuCGKP18}, probabilistic model repair \cite{bartocci2011model,chen2013model}, dynamic reconfiguration of self-adaptive systems \cite{Filieri2011,Filieri2013,10.1007/978-3-319-74183-3_1,calinescu2009cads}, and synthesis of confidence intervals for the QoS properties of software systems \cite{calinescu2016formal,calinescu2016fact}. These methods address very different problems, and yet most researchers who developed them mention the same limitation of parametric model checking: its computationally intensive nature. Addressing this one limitation can greatly improve the scalability and applicability of all the methods that use parametric model checking. Despite this significant incentive, research to improve PMC efficiency has been very limited so far. To the best of our knowledge, this research includes only the results from \cite{Hahn2011,Jansen2014}. As we explain in the rest of this section, these results represent significant advances, but are both complementary to our ePMC work.

The PMC technique presented in \cite{Hahn2011} provides major performance improvements over the initial PMC approach from \cite{Daws:2004:SPM:2102873.2102899}. While the language-theoretic PMC approach from \cite{Daws:2004:SPM:2102873.2102899} uses a regular expression to encode the probability that a PCTL path formula is satisfied, \cite{Hahn2011} computes a rational expression for the probability of reaching a set of parametric MC states, and mitigates the explosion in expression size relative to the number of MC states by exploiting algebraic symmetry and cancelation properties of rational functions. A further improvement introduced in \cite{Hahn2011} is the application of arithmetic operations during the state elimination stage of the PMC algorithm, to simplify the rational expression as it is calculated. The  technique is implemented by the parametric model checkers PARAM \cite{Hahn2010} and PRISM \cite{prism}, and shown to considerably reduce the complexity of PMC in \cite{Hahn2011}. Our work builds on the PMC technique from \cite{Hahn2011} (when using the parametric model checking functionality of PRISM in the second ePMC stage, cf.~Section~\ref{sect:tool}). Furthermore, ePMC complements the results from \cite{Hahn2011} by further speeding up parametric model checking through the pre-computation of PMC expressions for domain-specific modelling patterns.

The research from \cite{Jansen2014} introduces a compositional technique for parametric model checking. This technique decomposes the underlying state transition graph of the analysed MC into strongly connected components (SCCs). Rational functions are then computed independently for each SCC and then combined to obtain the PMC result. In addition, \cite{Jansen2014} defines new polynomial factorisations that further improve the handling of the large expressions generated by PMC, and optimises the computation of the greatest common divisor used to simplify PMC rational expressions. The PMC technique from \cite{Jansen2014} is implemented by the recently released probabilistic model checker Storm \cite{Dehnert2017}, and achieves significant performance improvements over the previously developed PMC techniques. Like the technique from \cite{Jansen2014}, ePMC is a compositional PMC method. However, while \cite{Jansen2014} operates with SCCs, the ePMC ``components'' are Markov chain fragments that may contain zero or more SCCs, or even parts of SCCs. This makes ePMC particularly flexible, and different from the technique from \cite{Jansen2014}. Further advantages of our method include the precomputation of the PMC expression associated with the Markov chain fragments, and the use of sets of formulae that include these PMC expressions without combining them. Finally, through using Storm in its second stage (cf.~Section~\ref{sect:tool}), ePMC leverages and extends the PMC technique from \cite{Jansen2014}. As shown in Section~\ref{sect:evaluation}, this significantly improves the efficiency and scalability of parametric model checking.

One other characteristic that distinguishes ePMC from the techniques in \cite{Daws:2004:SPM:2102873.2102899,Hahn2011,Jansen2014} is its use of a domain-specific repository of precomputed QoS property expressions. As such, our ePMC method and pattern-aware probabilistic model checker do not offer the generality of the other techniques and model checkers. In return---for the domains for which such a repository has been built---ePMC can analyse parametric Markov chains up to several orders of magnitude faster, and yields much smaller and much more efficient to evaluate formulae than the current PMC approaches.

\section{Conclusion \label{sect:conclusion}}

We presented ePMC, a tool-supported method for efficient parametric model checking. ePMC can efficiently analyse unbounded until and reachability reward PCTL formulae  by precomputing closed-form expressions for the QoS properties of modelling patterns used frequently within a domain of interest. These expressions are then employed to considerably speed up the analysis of Markov chain models of systems from the same domain, and to generate sets of QoS property formulae that can be evaluated very efficiently. These improvements extend the applicability of parametric model checking to much larger models than previously feasible. 

In our future work, we plan to extend the use of ePMC to further types of component-based systems. In particular, probabilistic model checking is increasingly used to analyse Markov chains comprising interchangeable modules within the important and broad domain of software product lines (e.g., \cite{DBLP:conf/splc/GhezziS11,DBLP:journals/infsof/GhezziS13,Chrszon2018,ter2018framework}). These interchangeable modules represent ideal ePMC modelling pattern candidates.

Furthermore, we envisage that the benefits of our work will extend to multiple applications of probabilistic and parametric model checking, and we plan to exploit ePMC in several of these applications. Thus, we intend to integrate ePMC with probabilistic model synthesis \cite{DBLP:conf/kbse/GerasimouTC15,DBLP:journals/ase/GerasimouCT18}, which is currently very computationally intensive due to the need to analyse numerous probabilistic model variants corresponding to different parameter values. Additionally, we plan to use ePMC instead of traditional parametric model checking in our recently introduced technique for formal verification with confidence intervals \cite{calinescu2016formal,calinescu2016fact}, which can only analyse QoS properties defined by small to medium size closed-form expressions. Last but not least, we will build on our recent work from \cite{Calinescu2017} to explore the use of ePMC within self-adaptive systems where not only the system parameters but also the system architecture needs to be reconfigured at runtime.

\section*{Acknowledgements}

This work was partly funded by the Assuring Autonomy International Programme.

\section*{Appendix A$\;\;$ Theorem proofs}

\begin{proof}[Proof of Theorem~1] 
Let $A,A'$ be the sets of paths that satisfy the PCTL formula $\Phi_1 \mathrm{U}\: \Phi_2$ for the MC $M(S,s_0,\mathbf{P},L)$ and for the MC $M'(S',s'_0,\mathbf{P}',L')$ induced by  $F(Z,z_0,Z_\mathsf{out})$:
\[
A\!=\!\{\pi\!\in\!\mathit{Paths}^M\!(s_0) \!\mid\! \exists i\!>\!0. (\pi(i)\!\models\! \Phi_2\,\wedge\,\forall j\!<\!i.\pi(j)\!\models\! \Phi_1)\}
\]
\[
A'\!\!=\!\{\pi'\!\!\in\!\mathit{Paths}^{M'}\!(s'_0) \!\mid\! \exists i\!>\!0. (\pi'\!(i)\!\models\! \Phi_2\,\wedge\,\forall j\!<\!i.\pi'\!(j)\!\models\! \Phi_1)\}
\]
According to the semantics of PCTL (cf.\ Section~\ref{sect:prelim-2}), we need to show that $\mathrm{Pr}_{s_0} (A)=\mathrm{Pr}'_{s'_0}(A')$, where $\mathrm{Pr}_{s_0}$ and $\mathrm{Pr}'_{s'_0}$ are probability measures defined over $\mathit{Paths}^M\!(s_0)$ and $\mathit{Paths}^{M'}\!(s'_0)$, respectively, as explained in Section~\ref{sect:prelim-1}. A path $\pi\in A$ has the general form
\begin{equation}
\label{eq:hybrid-path}
   \pi = \pi_0\omega_1\pi_1\omega_2\pi_2\ldots\omega_n\pi_n\ldots,
\end{equation}
where $\pi_0, \pi_1,\ldots,\pi_n$ are subpaths comprising only states from $S\setminus Z$, $\omega_1,\omega_2,\ldots,\omega_n$ are subpaths comprising only states from $Z$, and the first state of $\pi$ that satisfies $\Phi_2$ is the last state from the path prefix $\pi_0\omega_1\pi_1\omega_2\pi_2\ldots\omega_n\pi_n$. Note that~(\ref{eq:hybrid-path}) subsumes the scenarios when the path prefix starts with a state from $Z$ (subpath $\pi_0$ empty), ends with a state from $Z$ (subpath $\pi_n$ empty), contains only states from $S\setminus Z$ ($n=0$), or contains only states from $Z$ ($n=1$, and subpaths $\pi_0$ and $\pi_1$ empty). We have three cases.

\emph{Case 1)} If $n>0$ and $\pi_n$ is non-empty, the subpaths $\omega_1$, $\omega_2$, \ldots, $\omega_n$ must all start with $z_0$ and end with a state $z\in Z_\mathsf{out}$ (as paths can only ``enter'' and ``exit'' MC fragments through their only entry state and one of their output states, respectively). Furthermore, $z_0$ and $z$ must satisfy $\Phi_1$ (since it belongs to the prefix of path $\pi$), so all states from $Z$ also satisfy $\Phi_1$ (as required by the theorem). This means that substituting any subset of $\omega_1$, $\omega_2$, \ldots, $\omega_n$ from $\pi$ with any combination of fragment subpaths starting at $z_0$ and ending with a state from $Z_\mathsf{out}$ changes $\pi$ into a sequence of states that either belongs to $A$ (if it is a path from $\mathit{Paths}^M(s_0)$) or has probability 0 of occurring in $M$ (otherwise). Given the set of all paths $X(\pi_0, \pi_1,\ldots,\pi_n) \subseteq A$ that can be obtained through such substitutions, and using the notation $\pi_i^\mathsf{last}$ for the last state on subpath $\pi_i$, $1\leq i\leq n$, we have:
\begin{multline*}
   \!\!\!\!\!\!\mathrm{Pr}_{s_0} (X(\pi_0, \pi_1,\ldots,\pi_n)) = \\
   \!\left[\prod_{i=0}^{n-1}\! \mathbf{P}(\pi_i)\mathbf{P}(\pi_i^\mathsf{last},z_0)\!\!\!\sum_{z\in Z_\mathsf{out}}\!\!\! \mathit{prob}_z\mathbf{P}(z,\pi_{i+1}(1))\right]\!\mathbf{P}(\pi_n)\!=\\
\;\!\left[\prod_{i=0}^{n-1}\! \mathbf{P}'(\pi_i)\mathbf{P}'(\pi_i^\mathsf{last},\overline{z})\mathbf{P}'(\overline{z},\pi_{i+1}(1))\right]\!\mathbf{P}'(\pi_n)\!=\\
\mathrm{Pr}'_{s'_0} (X'(\pi_0, \pi_1,\ldots,\pi_n))
   \end{multline*}
where $X'(\pi_0, \pi_1,\ldots,\pi_n)=\{ \pi'\!\!\in\!\mathit{Paths}^{M'}\!(s'_0) \!\mid\! \pi'=\pi_0\overline{z}\pi_1$ $\overline{z}\pi_2\ldots\overline{z}\pi_n\ldots \}\subseteq A'$ (since all states $\pi_0, \pi_1,\ldots,\pi_n$ and $\overline{z}$ satisfy $\Phi_1$, and $\pi_n^\mathsf{last}$ satisfies $\Phi_2$). We can similarly show that any subset $X'(\pi_0, \pi_1,\ldots,\pi_n)$ of $A'$ with paths of the form $\pi_0\overline{z}\pi_1$ $\overline{z}\pi_2\ldots\overline{z}\pi_n\ldots$ and on which $\Phi_2$ first holds in state $\pi_n^\mathsf{last}$ corresponds to a subset $X(\pi_0, \pi_1,\ldots,\pi_n) \subseteq A$ such that $\mathrm{Pr}'_{s'_0} (X'(\pi_0, \pi_1,\ldots,\pi_n))=\mathrm{Pr}_{s_0} (X(\pi_0, \pi_1,\ldots,\pi_n))$.

\emph{Case 2)} If $n>0$ and $\pi_n$ is empty, then the last state of $\omega_n$ from~(\ref{eq:hybrid-path}) satisfies $\Phi_2$ and (since one state from $Z$ satisfies $\Phi_2$) all states from $Z$ must satisfy $\Phi_2$. This includes the states from $\omega_1$ and its initial state, $z_0$, so the set of paths $\pi$ form the set $X(\pi_0)\!=\! \{\pi\! \in\!\mathit{Paths}^M\!(s_0)\mid \pi=\pi_0z_0\ldots\}\subseteq A$, and we have: 
\begin{multline*}
  \mathrm{Pr}_{s_0}(X(\pi_0)) = \mathbf{P}(\pi_0)\mathbf{P}(\pi_0^\mathsf{last},z_0) = \\
  \mathbf{P}'(\pi_0)\mathbf{P}'(\pi_0^\mathsf{last},\overline{z}) = \mathrm{Pr}'_{s_0'}(X'(\pi_0)),
\end{multline*}
where $X'(\pi_0)=\{ \pi'\!\!\in\!\mathit{Paths}^{M'}\!(s'_0) \!\mid\! \pi'=\pi_0\overline{z}\ldots \}\subseteq A'$. In a similar way, we can show that any subset $A'(\pi_0)$ of $A'$ with paths of the form $\pi_o\overline{z}\ldots$ and on which $\Phi_2$ first holds in state $\overline{z}$ corresponds to a subset $X(\pi_0)$ of $A$ such that $\mathrm{Pr}'_{s'_0} (X'(\pi_0))=\mathrm{Pr}_{s_0} (X(\pi_0))$.

\emph{Case 3)} Finally, if $n=0$, the path $\pi$ from (\ref{eq:hybrid-path}) becomes $\pi=\pi_0\ldots$, and equiprobable sets $X(\pi_0)\subseteq A$ and $X'(\pi_0)\subseteq A'$ are straightforward to identify.

We have shown that $A$ and $A'$ can be partitioned into pairs of corresponding subsets $X\subseteq A$ and $X'\subseteq A'$ that are equiprobable according to the probability metrics $\mathrm{Pr}_{s_0}$ and $\mathrm{Pr}'_{s'_0}$, which completes the proof.
\end{proof}

\begin{proof}[Proof of Theorem~2] 
We adopt one of the alternative definitions for the reachability reward $\mathcal{R}_{=?} [\mathrm{F}\,T]$ from~\cite[\S 10.5.1]{BaierKatoen2008}. Given the set $A$ of all finite paths $\pi=s_0s_1\ldots s_m$ from $M$ such that $s_m\in T$ and $s_0,s_1,\ldots,s_{m-1}\notin T$, we have: 
\begin{equation}
\label{eq:th2-proof}
  \mathcal{R}_{=?} [\mathrm{F}\,T]=
  \left\{\begin{array}{ll}
  0, & \textrm{if $\mathcal{P}_{<1}[\mathrm{F}\, T]$}\\
  \sum_{\pi\in A} \mathbf{P}(\pi) \rho(\pi), & \textrm{otherwise}
  \end{array}\right.
\end{equation}
where $\rho(\pi) = \sum_{i=0}^{m-1} \rho(s_i)$.

According to Theorem~1, if $\mathcal{P}_{<1}[\mathrm{F}\, T]$  
over $M$ then $\mathcal{P}_{<1}[\mathrm{F}\, T]$ over $M'$ too, so $\mathcal{R}_{=?} [\mathrm{F}\,T]=0$ over both $M$ and $M'$ and the theorem holds. The theorem also holds trivially when $A$ contains no paths with states from $Z$: in this case, it is straightforward to show that $A$ is also the set of all finite paths $s_0s_1\ldots s_m$ from $M'$ such that $s_m\in T$ and $s_0,s_1,\ldots,s_{m-1}\notin T$. 
As such, the rest of the proof considers the case when $\mathcal{P}_{\geq 1}[\mathrm{F}\, T]$ and $A$ contains paths with states from $Z$. The first state from $Z$ on such a path will be $z_0$ (the only entry state of the fragment $F$), immediately followed by other states from $Z$ until a state from $Z_\mathsf{out}$ is reached, and then followed by at least one state from $S\setminus Z$ (since $T\cap Z=\emptyset$ and $Z_\mathsf{out}$ states are followed by a state from outside $Z$). Thus, the generic form of such a path is 
\begin{equation}
\label{eq:path-theorem-2}
  \pi=\underbrace{\overbrace{s_0\ldots s_i}^{\in S\setminus Z}\overbrace{s_{i+1}}^{=z_0}\overbrace{s_{i+2}\ldots s_{j-1}}^{\in Z}\overbrace{s_j}^{\in Z_\mathsf{out}}\overbrace{s_{j+1}}^{\in S\setminus Z}\overbrace{s_{j+2}\ldots s_{m-1}}^{\in S}}_{\notin T}\underbrace{s_{m}}_{\in T}
\end{equation}
We consider the subset of all paths $A_1\subseteq A$ that start with the prefix $s_0\ldots s_iz_0$ and, using the notation $\pi_{x,y}=s_xs_{x+1}\ldots$ $s_y$, we calculate their contribution to the sum from the second row of (\ref{eq:th2-proof}):
\[
\!\!\!\begin{array}{l}
C(s_0,s_1,\ldots,s_i,z_0) = \sum_{\pi\in A_1} \mathbf{P}(\pi) \rho(\pi)=\\ 
\sum_{\pi\in A_1} \mathbf{P}(\pi_{0,i+1})\mathbf{P}(\pi_{i+1,m})(\rho(\pi_{0,i+1})\!+\!\rho(\pi_{i+1,m}))=\\
\mathbf{P}(\pi_{0,i+1})\left[\left(\sum_{\pi\in A_1} \mathbf{P}(\pi_{i+1,m})\right)\rho(\pi_{0,i+1}) + \right.\\
\qquad\qquad\qquad\qquad\;\;\;\left.\sum_{\pi\in A_1} \mathbf{P}(\pi_{i+1,m})\rho(\pi_{i+1,m})\right]=\\
\mathbf{P}(\pi_{0,i+1})\left[ 1\cdot \rho(\pi_{0,i+1}) + 
\sum_{\pi\in A_1}\!\! \mathbf{P}(\pi_{i+1,m})\rho(\pi_{i+1,m})\right],
\end{array}
\]
since $\mathcal{P}_{\geq 1}[\mathrm{F}\, T]$ requires that $\sum_{\pi\in A_1} \mathbf{P}(\pi_{i+1,m})=1$. Additionally, using the shorthand notation $\pi_{z_0\rightarrow z}$ and $\pi_{j+1\rightarrow T}$ for the set of all sub-paths $s_{i+1}s_{i+2}\ldots s_{j}$ associated with a fixed $s_j=z\in Z_\mathsf{out}$ and for the set of all sub-paths $s_{j+1}s_{j+2}\ldots s_{m}$ from~(\ref{eq:path-theorem-2}), respectively, we have:
\[
\!\!\!\begin{array}{l}
\sum_{\pi\in A_1}\!\! \mathbf{P}(\pi_{i+1,m})\rho(\pi_{i+1,m})=\\
\sum_{z\in Z_\mathsf{out}} \sum_{\pi_1\in \pi_{z_0\rightarrow z}}\sum_{\pi_2\in\pi_{j+1\rightarrow T}} \mathbf{P}(\pi_1)\mathbf{P}(z,s_{j+1})\mathbf{P}(\pi_2)\\
\qquad\qquad\qquad\qquad\qquad\qquad\qquad\;\;\;(\rho(\pi_1)+\rho(z)+\rho(\pi_2))=\\
\sum_{z\in Z_\mathsf{out}} \sum_{\pi_1\in \pi_{z_0\rightarrow z}} \bigl[\mathbf{P}(\pi_1)(\rho(\pi_1) +\rho(z))\cdot\\
\qquad\qquad\qquad\qquad\quad\sum_{\pi_2\in\pi_{j+1\rightarrow T}} \mathbf{P}(z,s_{j+1})\mathbf{P}(\pi_2)\bigr] +\\
\sum_{z\in Z_\mathsf{out}} \bigl[\bigl(\sum_{\pi_1\in \pi_{z_0\rightarrow z}}\mathbf{P}(\pi_1)\bigr)\cdot \\
\qquad\qquad\qquad\qquad\quad\sum_{\pi_2\in\pi_{j+1\rightarrow T}} \mathbf{P}(z,s_{j+1}) \mathbf{P}(\pi_2)\rho(\pi_2))\bigr]=\\
\sum_{z\in Z_\mathsf{out}} \sum_{\pi_1\in \pi_{z_0\rightarrow z}}\bigl[\mathbf{P}(\pi_1)(\rho(\pi_1)+\rho(z))\cdot 1\bigr] +\\
\sum_{z\in Z_\mathsf{out}} \bigl[\,\mathit{prob}_z\cdot \sum_{\pi_2\in\pi_{j+1\rightarrow T}} \mathbf{P}(z,s_{j+1}) \mathbf{P}(\pi_2)\rho(\pi_2)\bigr]=\\
\overline{\mathit{rwd}} +\sum_{\pi_2\in\pi_{j+1\rightarrow T}}  \left(\sum_{z\in Z_\mathsf{out}} \mathit{prob}_z\mathbf{P}(z,s_{j+1})\right) \mathbf{P}(\pi_2)\rho(\pi_2)=\\
\overline{\mathit{rwd}} + \sum_{\pi_2\in\pi_{j+1\rightarrow T}} \mathbf{P}'(\overline{z},s_{j+1})\mathbf{P}(\pi_2)\rho(\pi_2),
\end{array}
\]
where $\overline{\mathit{rwd}}$, $\mathit{prob}_z$ and $\mathbf{P}'(\overline{z},s_{j+1})$ are those from Definition~\ref{def:abstractMC}, and where we used the fact that $\mathcal{P}_{\geq 1}[\mathrm{F}\, T]$ 
to infer that $\sum_{\pi_2\in\pi_{j+1\rightarrow T}} \mathbf{P}(z,s_{j+1})\mathbf{P}(\pi_2)=1$ and to include all non-zero-probability sub-paths at every step of the calculation. Combining the results so far, we obtain
\[
\!\!\!\begin{array}{l}
  C(s_0,s_1,\ldots,s_i,z_0)\!=\!\mathbf{P}(\pi_{0,i+1})\left(\rho(\pi_{0,i+1})+\overline{\mathit{rwd}} \;+ \right.\\
  \qquad\qquad\qquad\qquad\qquad\!\!\;\sum_{\pi_2\in\pi_{j+1\rightarrow T}} \mathbf{P}'(\overline{z},s_{j+1})\mathbf{P}(\pi_2)\rho(\pi_2)).
  \end{array}
  \]
This reward ``contribution'' is equivalent to that obtained by replacing all sub-paths from $F$ appearing in paths from $A_1$ immediately after the prefix $s_0s_1\ldots s_i$ with state $\overline{z}$ comprising a reward value $\overline{\mathit{rwd}}$ and transition probabilities $\mathbf{P}'(\overline{z},s_{j+1})$ to states $s_{j+1}\in S\setminus Z$. By repeating this  process to replace all occurrences of sub-paths from $F$ appearing in $A$, we obtain an expression equivalent to $\sum_{\pi\in A} \mathbf{P}(\pi) \rho(\pi)$, but corresponding to the parametric model checking of $\mathcal{R}_{=?} [\mathrm{F}\,T]$ over the abstract MC $M'$ induced by $F$, which completes the proof.
\end{proof}

\begin{proof}[Proof of Theorem~3]
We prove the $\mathsf{SEQ}$ results by induction. For the base case, we have $n=1$, corresponding to an SBS operation carried out by a single service with success probability $p_1$, cost $c_1$ and execution time $t_1$. As required, $p_\mathsf{SEQ}=p_1=1-(1-p_1)$, $c_\mathsf{SEQ}=c_1$ and $t_\mathsf{SEQ}=t_1$. Assume now that the $\mathsf{SEQ}$ expressions from Table~\ref{tab:sbs_expressions} are correct for $n$ services, and consider an $\mathsf{SEQ}$ pattern comprising $n+1$ services. There are two ways in which the $n+1$ services can complete the operation successfully: (i)~either the first $n$ services complete the operation successfully (with probability $1-\prod_{i=1}^n (1-p_i)$), or (ii)~each of the first $n$ services fails, and the invocation of the $(n+1)$-th service is successful. Accordingly, the success probability for the $(n+1)$-service $\mathsf{SEQ}$ pattern is:
\[
\begin{array}{l}
  \!\!\left(1\!-\!\prod_{i=1}^n (1\!-\!p_i)\right) \!+\! \left(\left(\prod_{i=1}^n (1\!-\!p_i)\right) p_{n+1}\right) \!\!=\! 1\!-\!\prod_{i=1}^{n+1} (1\!-\!p_i).
\end{array}
\]
To calculate the expected cost and execution time for the $(n+1)$-service $\mathsf{SEQ}$ pattern, recall that the $(n+1)$-th service is invoked iff the invocations of all previous $n$ services failed, i.e.\ with probability $\prod_{i=1}^n (1-p_i)$. Accordingly, using the $(n+1)$-th service adds a supplementary expected cost of $\left(\prod_{i=1}^n (1-p_i)\right)c_{n+1}$ and a supplementary expected execution time of $\left(\prod_{i=1}^n (1-p_i)\right)t_{n+1}$ to the expected cost and execution time of an $n$-service $\mathsf{SEQ}$ pattern, respectively.  Thus, the expected cost for the $(n+1)$-service $\mathsf{SEQ}$ pattern is given by
\[
\begin{array}{l}
  \left(\!c_1\!+\!\sum_{i=2}^n \left(\prod_{j=1}^{i-1}(1\!-\!p_j)\right)c_i\right) + \left(\prod_{i=1}^n (1\!-\!p_i)\right)c_{n+1} = \\
  \qquad\qquad\qquad\qquad\qquad\qquad = c_1+\sum_{i=2}^{n+1} \left(\prod_{j=1}^{i-1}(1\!-\!p_j)\right)c_i
\end{array}
\]
and the expected execution time can be calculated similarly, which completes the induction step.

For the $\mathsf{PAR}$ pattern, the probability that the parallel invocations of the $n$ (independent) services will all fail is $\prod_{i=1}^n (1-p_i)$, so the probability that the operation will be completed successfully is
$1-\prod_{i=1}^n (1-p_i)$ as required. Also, since all $n$ services are always invoked, the cost for the pattern is given by the sum of the $n$ service costs. Finally, to calculate the expected execution time for the $\mathsf{PAR}$ pattern, assume (as stated in Table~\ref{tab:sbs_expressions} and without loss of generality) that the $n$ services are ordered such that $t_1\leq t_2\leq\cdots\leq t_n$. Under this assumption, service $i$ will be the first service that completes execution \emph{successfully} (in time $t_i$) iff: (i)~the invocations of the faster services $1,2,\ldots,i-1$ have all failed (which happens with probability $\prod_{j=1}^{i-1}(1-p_j)$); and (ii)~the invocation of service $i$ is successful (which happens with probability $p_i$). Thus, the execution time for the $\mathsf{PAR}$ patterns follows a discrete distribution with probability $\left(\prod_{j=1}^{i-1}(1-p_j)\right)p_i$ of successful completion in time $t_i$, and probability $\prod_{j=1}^{n}(1-p_j)$ of unsuccessful completion in time $t_n$. As a result, the expected execution time for the pattern is given by:
\[
\begin{array}{l}
   \sum_{i=1}^n \left(\prod_{j=1}^{i-1}(1-p_j)\right)p_it_i + \left(\prod_{j=1}^{n}(1-p_j)\right)t_n =  p_1t_1 + \\ 
   \;\; + \sum_{i=2}^{n-1} \left(\prod_{j=1}^{i-1}(1-p_j)\right)p_it_i + \left[\left(\prod_{j=1}^{n-1}(1-p_j)\right)p_nt_n\right. + \\
   \;\; +\left.\left(\prod_{j=1}^{n-1}(1-p_j)\right)(1-p_n)t_n\right] = \\
   \;\;= p_1t_1 \!+\! \sum_{i=2}^{n-1} \left(\prod_{j=1}^{i-1}(1-p_j)\right)p_it_i \!+\! \left(\prod_{j=1}^{n-1}(1-p_j)\right)t_n,
   \end{array}
\]
which can be easily rearranged in the format from Table~\ref{tab:sbs_expressions} by introducing the notation $\widetilde{p}_i=p_i$ for $i<n$ and $\widetilde{p}_n=1$.

For the $\mathsf{PROB}$ pattern, the results from Table~\ref{tab:sbs_expressions} follow immediately from the fact that the success probability, cost and execution time of the operation have a discrete distribution with probabilities $x_1$, $x_2$, \ldots, $x_n$ of taking the values $p_1$, $p_2$, \ldots, $p_n$ (for the success probability), $c_1$, $c_2$, \ldots, $c_n$ (for the cost), and $t_!$, $t_2$, \ldots, $t_n$ (for the execution time).

For the $\textsf{SEQ\_R1}$ pattern, we first focus on a single service $i$ with success probability $p_i$, cost $c_i$ and execution time $t_i$. If unsuccessful invocations of the service (which happen with probability $(1-p_i)$) are followed by its re-invocation with probability $r_i$, then the overall probability of successfully invoking the service is given by:
\begin{equation}
\label{eq:1}
\begin{array}{lll}
   p'_i&\!\!\!\!=\!\!\!\!&p_i + (1-p_i)r_i \biggl(p_i + (1-p_i)r_i\bigl(p_i+\ldots \bigr)\biggr) =\\
   &\!\!\!\!=\!\!\!\!& p_i + [(1-p_i)r_i]p_i + [(1-p_i)r_i]^2p_i + \ldots =\\[1.5mm]
   &\!\!\!\!=\!\!\!\!& \lim_{m\rightarrow\infty} \sum_{j=0}^m [(1-p_i)r_i]^jp_i =\\[1.5mm]
   &\!\!\!\!=\!\!\!\!& \lim_{m\rightarrow\infty} p_i\frac{1-[(1-p_i)r_i]^{m+1}}{1-[(1-p_i)r_i]} = \frac{p_i}{1-(1-p_i)r_i}.
   \end{array}
\end{equation}
A similar reasoning can be used to show that the expected cost and execution time of the service with re-invocations are 
\begin{equation}
\label{eq:2}
  c'_i=\frac{c_i}{1-(1-p_i)r_i} \;\;\textrm{ and }\;\; t'_i=\frac{t_i}{1-(1-p_i)r_i}, 
\end{equation}
respectively. 
Thus, the $n$-service $\mathsf{SEQ\_R1}$ pattern from Table~\ref{tab:sbs_expressions} is equivalent to an $n$-service $\mathsf{SEQ}$ pattern whose services have success probabilities $p'_1$, $p'_2$, \ldots, $p'_n$, costs $c'_1$, $c'_2$, \ldots, $c'_n$, and execution times $t'_1$, $t'_2$, \ldots, $t'_n$. As a result, the expressions for the success probability, expected cost and expected execution time of the $\mathsf{SEQ\_R1}$ pattern can be obtained by using these parameters in the analogous expressions of the $\mathsf{SEQ}$ pattern, as shown in Table~\ref{tab:sbs_expressions}.

The $\mathsf{SEQ\_R}$ pattern is equivalent to having a single service with success probability $p_\mathsf{SEQ}$, cost  $c_\mathsf{SEQ}$ and execution time $t_\mathsf{SEQ}$, and re-invoking this service with probability $r$ after unsuccessful invocations. As such, the success probability, expected cost and expected execution time for the $\mathsf{SEQ\_R}$ pattern are obtained by applying the formulae from~(\ref{eq:1}) and~(\ref{eq:2}) to this equivalent service, which yields the expressions from Table~\ref{tab:sbs_expressions}.

Using the same reasoning as for the $\mathsf{SEQ\_R}$ pattern, it is straightforward to show that Table~\ref{tab:sbs_expressions} provides the correct expressions for the $\mathsf{PAR\_R}$ and $\mathsf{PROB\_R}$ patterns.

Finally, the $n$-service $\mathsf{PROB\_R1}$ pattern is equivalent to an $n$-service $\mathsf{PROB}$ pattern whose $i$-th service has success probability $p'_i$ given by~(\ref{eq:1}), and cost $c'_i$ and execution time $t'_i$ given by~(\ref{eq:2}). Using these three formulae as parameters in the expressions giving the success probability, expected cost and expected execution time of the $\mathsf{PROB}$ pattern yields the results for the $\mathsf{PROB\_R1}$ pattern.
\end{proof}

\begin{proof}[Proof of Theorem~4]
For the \textsf{BASIC} pattern, either the server remains operational and all $n_1, n_2,\ldots,n_m$ tier instances are still running at the end of the analysed time period, or the server fails and no instance is left running. The former scenario occurs with probability $p$, so $p_{b_1,b_2,\ldots,b_m}=p$ iff $b_i=1$ for all tiers $i$ for which $n_i=1$ and $b_i=2+$ for all tiers $i$ for which $n_i>1$; and the latter scenario occurs with probability $1-p$, so $p_{0,0,\ldots,0}=1-p$. Otherwise, $p_{b_1,b_2,\ldots,b_m}=0$ since there is no scenario in which only some of the tier instances are left running and others are lost.

For the \textsf{VIRTUALIZED} pattern, we first consider the scenario where at least one of $b_1$, $b_2$, \ldots, $b_m$ is non-zero. This requires that $m+1$ independent events occur: server stays up; and the appropriate number of VMs running instances of tier $i\in\{1,2,\ldots,m\}$ (i.e.\ zero, one, or greater than one) remain operational. The probability of the first event is $p$, and the probability of each of the other events is given by the probability that the value of a random variable with binomial distribution $\mathcal{B}(n_i,p_\mathsf{VM})$ is: zero (i.e.\ $(1-p_\mathsf{VM})^{n_i}=f(0,n_i)$); one (i.e.\ $n_ip_\mathsf{VM}(1-p_\mathsf{VM})^{n_i-1}=f(1,n_i)$); or greater than one (i.e.\ $1-f(0,n_i)-f(1,n_i)=f(2+,n_i)$). The first part of the result from Table~\ref{tab:server_expressions} is obtained by multiplying these $m+1$ probabilities.  Finally, the scenario $b_1=b_2=\cdots=b_m=0$ occurs in two circumstances: when the server fails---which happens with probability $(1-p)$, and when the server stays up but all $n_1+n_2+\cdots n_m$ VMs running tier instances fail---which happens with probability $p(1-p_\mathsf{VM})^{\sum_{i=1}^m n_i}$, giving the last part of the result for the $\mathsf{VIRTUALIZED}$ pattern.

Finally, for the \textsf{VIRTUALIZED-M} pattern, the scenario where at least one of $b_1$, $b_2$, \ldots, $b_m$ is non-zero can occur in two cases. In the first case, the server stays up and the appropriate number of VMs from each tier remains operational, which has probability $p\prod_{i=1}^m f(b_i,n_i)$ (as shown above for the $\mathsf{VIRTUALIZED}$ pattern); this corresponds to the first term from the definition of $p_{b_1,b_2,\ldots,b_m}$ from Table~\ref{tab:server_expressions}. In the second case, the server fails but the failure is detected (which happens with probability $(1-p)p_\mathsf{detect}$), and $b_i$ VMs running tier $i$ are successfully migrated to other servers and remain operational for $i=1,2,\ldots,m$. To prove that the second term from the definition of $p_{b_1,b_2,\ldots,b_m}$ is correct, we will show that the probability of this last event is $g(b_i,n_i)$ from Table~\ref{tab:server_expressions} for all three values of $b_i$. We start by noting that a given VM from tier $i$ is successfully migrated with probability $\frac{p_\mathsf{migrate}}{1-(1-p_\mathsf{migrate}r)}$, a result that can be obtained as in~(\ref{eq:1}); so the VM will be migrated \emph{and} remain operational with probability  $\frac{p_\mathsf{migrate}p_\mathsf{VM}}{1-(1-p_\mathsf{migrate}r)}$. Accordingly, $g(0,n_i)$, $g(1,n_i)$ and $g(2+,n_i)$ represent the probabilities that a random variable with binomial distribution $\mathcal{B}\!\left(n_i,\frac{p_\mathsf{migrate}p_\mathsf{VM}}{1-(1-p_\mathsf{migrate}r)}\right)$ takes values $0$, $1$, or greater than or equal to $2$, respectively, which corresponds to the definition of $g(b_i,n_i)$ from Table~\ref{tab:server_expressions}. We complete the proof by noting that the scenario where $b_1=b_2=\cdots=b_m=0$ occurs in the same two cases, as well as when the server fails and its failure is not detected (which happens with probability $(1-p)(1-p_\mathsf{detect})$, and corresponds to the last term from the definition of $p_{b_1,b_2,\ldots,b_m}$ for this scenario).
\end{proof}



\begin{thebibliography}{10}
\providecommand{\url}[1]{#1}
\csname url@samestyle\endcsname
\providecommand{\newblock}{\relax}
\providecommand{\bibinfo}[2]{#2}
\providecommand{\BIBentrySTDinterwordspacing}{\spaceskip=0pt\relax}
\providecommand{\BIBentryALTinterwordstretchfactor}{4}
\providecommand{\BIBentryALTinterwordspacing}{\spaceskip=\fontdimen2\font plus
\BIBentryALTinterwordstretchfactor\fontdimen3\font minus
  \fontdimen4\font\relax}
\providecommand{\BIBforeignlanguage}[2]{{%
\expandafter\ifx\csname l@#1\endcsname\relax
\typeout{** WARNING: IEEEtranS.bst: No hyphenation pattern has been}%
\typeout{** loaded for the language `#1'. Using the pattern for}%
\typeout{** the default language instead.}%
\else
\language=\csname l@#1\endcsname
\fi
#2}}
\providecommand{\BIBdecl}{\relax}
\BIBdecl

\bibitem{Andova2004}
S.~Andova, H.~Hermanns, and J.-P. Katoen, ``Discrete-time rewards
  model-checked,'' in \emph{First International Workshop on Formal Modeling and
  Analysis of Timed Systems (FORMATS)}, 2004, pp. 88--104.

\bibitem{BaierKatoen2008}
C.~Baier and J.-P. Katoen, \emph{Principles of Model Checking}.\hskip 1em plus
  0.5em minus 0.4em\relax MIT Press, 2008.

\bibitem{bartocci2011model}
E.~Bartocci, R.~Grosu, P.~Katsaros, C.~Ramakrishnan, and S.~Smolka, ``Model
  repair for probabilistic systems,'' in \emph{17th International Conference on
  Tools and Algorithms for the Construction and Analysis of Systems (TACAS)},
  2011, pp. 326--340.

\bibitem{BenediktLW2013}
M.~Benedikt, R.~Lenhardt, and J.~Worrell, ``{LTL} model checking of interval
  {Markov} chains,'' in \emph{19th International Conference on Tools and
  Algorithms for the Construction and Analysis of Systems (TACAS)}, 2013, pp.
  32--46.

\bibitem{5611553}
R.~Calinescu, L.~Grunske, M.~Kwiatkowska, R.~Mirandola, and G.~Tamburrelli,
  ``Dynamic {QoS} management and optimization in service-based systems,''
  \emph{IEEE Transactions on Software Engineering}, vol.~37, no.~3, pp.
  387--409, May 2011.

\bibitem{6693145}
R.~Calinescu, K.~Johnson, and Y.~Rafiq, ``Developing self-verifying
  service-based systems,'' in \emph{28th IEEE/ACM International Conference on
  Automated Software Engineering (ASE)}, 2013, pp. 734--737.

\bibitem{Calinescu2017}
R.~Calinescu, D.~Weyns, S.~Gerasimou, M.~U. Iftikhar, I.~Habli, and T.~Kelly,
  ``Engineering trustworthy self-adaptive software with dynamic assurance
  cases,'' \emph{IEEE Transactions on Software Engineering}, vol.~44, no.~11,
  pp. 1039--1069, 2018.

\bibitem{calinescu2017synthesis}
R.~Calinescu, M.~Autili, J.~C{\'a}mara, A.~Di~Marco, S.~Gerasimou,
  P.~Inverardi, A.~Perucci, N.~Jansen, J.-P. Katoen, M.~Kwiatkowska
  \emph{et~al.}, ``Synthesis and verification of self-aware computing
  systems,'' in \emph{Self-Aware Computing Systems}.\hskip 1em plus 0.5em minus
  0.4em\relax Springer, 2017, pp. 337--373.

\bibitem{DBLP:journals/jss/CalinescuCGKP18}
R.~Calinescu, M.~{\v{C}}e{\v{s}}ka, S.~Gerasimou, M.~Kwiatkowska, and
  N.~Paoletti, ``Efficient synthesis of robust models for stochastic systems,''
  \emph{Journal of Systems and Software}, vol. 143, pp. 140--158, 2018.

\bibitem{calinescu2016formal}
R.~Calinescu, C.~Ghezzi, K.~Johnson, M.~Pezz{\'e}, Y.~Rafiq, and
  G.~Tamburrelli, ``Formal verification with confidence intervals to establish
  quality of service properties of software systems,'' \emph{IEEE Transactions
  on Reliability}, vol.~65, no.~1, pp. 107--125, 2016.

\bibitem{calinescu2016fact}
R.~Calinescu, K.~Johnson, and C.~Paterson, ``{FACT: A} probabilistic model
  checker for formal verification with confidence intervals,'' in \emph{22nd
  International Conference on Tools and Algorithms for the Construction and
  Analysis of Systems (TACAS)}, 2016, pp. 540--546.

\bibitem{DBLP:conf/icse/CalinescuJP18}
------, ``Efficient parametric model checking using domain-specific modelling
  patterns,'' in \emph{40th International Conference on Software Engineering:
  New Ideas and Emerging Results (ICSE:NIER)}, 2018, pp. 61--64.

\bibitem{calinescu2012compositional}
R.~Calinescu, S.~Kikuchi, and K.~Johnson, ``Compositional reverification of
  probabilistic safety properties for large-scale complex {IT} systems.'' in
  \emph{17th Monterey Workshop: Large-Scale Complex IT Systems}, 2012, pp.
  303--329.

\bibitem{calinescu2009cads}
R.~Calinescu and M.~Kwiatkowska, ``{CADS*}: {C}omputer-aided development of
  self-* systems,'' in \emph{12th International Conference on Fundamental
  Approaches to Software Engineering (FASE)}.\hskip 1em plus 0.5em minus
  0.4em\relax Springer, 2009, pp. 421--424.

\bibitem{CARDOSO2004}
J.~Cardoso, A.~Sheth, J.~Miller, J.~Arnold, and K.~Kochut, ``Quality of service
  for workflows and web service processes,'' \emph{Web Semantics: Science,
  Services and Agents on the World Wide Web}, vol.~1, no.~3, pp. 281--308,
  2004.

\bibitem{chen2013model}
T.~Chen, E.~M. Hahn, T.~Han, M.~Kwiatkowska, H.~Qu, and L.~Zhang, ``Model
  repair for {M}arkov decision processes,'' in \emph{International Symposium on
  Theoretical Aspects of Software Engineering (TASE)}, 2013, pp. 85--92.

\bibitem{Chrszon2018}
P.~Chrszon, C.~Dubslaff, S.~Kl{\"u}ppelholz, and C.~Baier, ``{ProFeat}:
  feature-oriented engineering for family-based probabilistic model checking,''
  \emph{Formal Aspects of Computing}, vol.~30, no.~1, pp. 45--75, 2018.

\bibitem{Ciesinski2004}
F.~Ciesinski and M.~Gr{\"o}{\ss}er, ``On probabilistic computation tree
  logic,'' in \emph{Validation of Stochastic Systems: A Guide to Current
  Research}, C.~Baier, B.~R. Haverkort, H.~Hermanns, J.-P. Katoen, and
  M.~Siegle, Eds.\hskip 1em plus 0.5em minus 0.4em\relax Springer, 2004, pp.
  147--188.

\bibitem{7968147}
C.~E. da~Silva, J.~D.~S. da~Silva, C.~Paterson, and R.~Calinescu,
  ``Self-adaptive role-based access control for business processes,'' in
  \emph{12th IEEE/ACM International Symposium on Software Engineering for
  Adaptive and Self-Managing Systems (SEAMS)}, 2017, pp. 193--203.

\bibitem{Daws:2004:SPM:2102873.2102899}
C.~Daws, ``Symbolic and parametric model checking of discrete-time {M}arkov
  chains,'' in \emph{First International Conference on Theoretical Aspects of
  Computing (ICTAC)}, 2005, pp. 280--294.

\bibitem{10.1007/978-3-319-74183-3_1}
R.~de~Lemos, D.~Garlan, C.~Ghezzi, H.~Giese, J.~Andersson, M.~Litoiu,
  B.~Schmerl, D.~Weyns, L.~Baresi, N.~Bencomo, Y.~Brun, J.~Camara,
  R.~Calinescu, M.~B. Cohen, A.~Gorla, V.~Grassi, L.~Grunske, P.~Inverardi,
  J.-M. Jezequel, S.~Malek, R.~Mirandola, M.~Mori, H.~A. M{\"u}ller, R.~Rouvoy,
  C.~M.~F. Rubira, E.~Rutten, M.~Shaw, G.~Tamburrelli, G.~Tamura, N.~M.
  Villegas, T.~Vogel, and F.~Zambonelli, ``Software engineering for
  self-adaptive systems: Research challenges in the provision of assurances,''
  in \emph{Software Engineering for Self-Adaptive Systems III. Assurances},
  R.~de~Lemos, D.~Garlan, C.~Ghezzi, and H.~Giese, Eds.\hskip 1em plus 0.5em
  minus 0.4em\relax Springer, 2017, pp. 3--30.

\bibitem{Dehnert2015}
C.~Dehnert, S.~Junges, N.~Jansen, F.~Corzilius, M.~Volk, H.~Bruintjes, J.-P.
  Katoen, and E.~{\'A}brah{\'a}m, ``{PROPhESY: A PRObabilistic ParamEter
  SYnthesis Tool},'' in \emph{27th International Conference on Computer Aided
  Verification (CAV)}, 2015, pp. 214--231.

\bibitem{Dehnert2017}
C.~Dehnert, S.~Junges, J.-P. Katoen, and M.~Volk, ``A {S}torm is coming: {A}
  modern probabilistic model checker,'' in \emph{29th International Conference
  Computer Aided Verification (CAV)}, 2017, pp. 592--600.

\bibitem{Filieri2011}
A.~Filieri, C.~Ghezzi, and G.~Tamburrelli, ``Run-time efficient probabilistic
  model checking,'' in \emph{33rd International Conference on Software
  Engineering (ICSE)}, 2011, pp. 341--350.

\bibitem{Filieri2013}
A.~Filieri and G.~Tamburrelli, ``Probabilistic verification at runtime for
  self-adaptive systems.'' \emph{Assurances for Self-Adaptive Systems}, vol.
  7740, pp. 30--59, 2013.

\bibitem{DBLP:journals/tse/FilieriTG16}
A.~Filieri, G.~Tamburrelli, and C.~Ghezzi, ``Supporting self-adaptation via
  quantitative verification and sensitivity analysis at run time,''
  \emph{{IEEE} Transactions on Software Engineering}, vol.~42, no.~1, pp.
  75--99, 2016.

\bibitem{Gallotti:2008:QPS:1478067.1478078}
S.~Gallotti, C.~Ghezzi, R.~Mirandola, and G.~Tamburrelli, ``Quality prediction
  of service compositions through probabilistic model checking,'' in \emph{4th
  International Conference on Quality of Software-Architectures (QoSA)}, 2008,
  pp. 119--134.

\bibitem{DBLP:journals/ase/GerasimouCT18}
S.~Gerasimou, R.~Calinescu, and G.~Tamburrelli, ``Synthesis of probabilistic
  models for quality-of-service software engineering,'' \emph{Automated
  Software Engineering Journal}, vol.~25, no.~4, pp. 785--831, 2018.

\bibitem{DBLP:conf/kbse/GerasimouTC15}
S.~Gerasimou, G.~Tamburrelli, and R.~Calinescu, ``Search-based synthesis of
  probabilistic models for quality-of-service software engineering,'' in
  \emph{30th {IEEE/ACM} International Conference on Automated Software
  Engineering (ASE)}, 2015, pp. 319--330.

\bibitem{DBLP:conf/splc/GhezziS11}
C.~Ghezzi and A.~M. Sharifloo, ``Verifying non-functional properties of
  software product lines: {T}owards an efficient approach using parametric
  model checking,'' in \emph{15th International Conference on Software Product
  Lines {SPLC}}, 2011, pp. 170--174.

\bibitem{DBLP:journals/infsof/GhezziS13}
------, ``Model-based verification of quantitative non-functional properties
  for software product lines,'' \emph{Information {\&} Software Technology},
  vol.~55, no.~3, pp. 508--524, 2013.

\bibitem{hahn2011synthesis}
E.~M. Hahn, T.~Han, and L.~Zhang, ``Synthesis for pctl in parametric markov
  decision processes,'' in \emph{NASA Formal Methods Symposium}.\hskip 1em plus
  0.5em minus 0.4em\relax Springer, 2011, pp. 146--161.

\bibitem{Hahn2010}
E.~M. Hahn, H.~Hermanns, B.~Wachter, and L.~Zhang, ``{PARAM: A} model checker
  for parametric {M}arkov models,'' in \emph{22nd International Conference on
  Computer Aided Verification (CAV)}, 2010, pp. 660--664.

\bibitem{Hahn2011}
E.~M. Hahn, H.~Hermanns, and L.~Zhang, ``Probabilistic reachability for
  parametric {M}arkov models,'' \emph{International Journal on Software Tools
  for Technology Transfer}, vol.~13, no.~1, pp. 3--19, 2011.

\bibitem{Hansson1994}
H.~Hansson and B.~Jonsson, ``A logic for reasoning about time and
  reliability,'' \emph{Formal Aspects of Computing}, vol.~6, no.~5, pp.
  512--535, 1994.

\bibitem{Jansen2014}
N.~Jansen, F.~Corzilius, M.~Volk, R.~Wimmer, E.~{\'A}brah{\'a}m, J.-P. Katoen,
  and B.~Becker, ``Accelerating parametric probabilistic verification,'' in
  \emph{11th International Conference on Quantitative Evaluation of Systems
  (QEST)}, 2014, pp. 404--420.

\bibitem{DBLP:conf/cbse/JohnsonCK13}
K.~Johnson, R.~Calinescu, and S.~Kikuchi, ``An incremental verification
  framework for component-based software systems,'' in \emph{16th {ACM}
  {SIGSOFT} Symposium on Component Based Software Engineering (CBSE)}, 2013,
  pp. 33--42.

\bibitem{Katoen:2011:IOP:1930080.1930191}
J.-P. Katoen, I.~S. Zapreev, E.~M. Hahn, H.~Hermanns, and D.~N. Jansen, ``The
  ins and outs of the probabilistic model checker {MRMC},'' \emph{Performance
  Evaluation}, vol.~68, no.~2, pp. 90--104, 2011.

\bibitem{kemeny-etal1976}
J.~G. Kemeny, J.~L. Snell, and A.~W. Knapp, \emph{Denumerable {M}arkov Chains,
  2nd edition}, ser. Graduate Texts in Marhematics.\hskip 1em plus 0.5em minus
  0.4em\relax Springer, 1976, vol.~40.

\bibitem{prism}
M.~Kwiatkowska, G.~Norman, and D.~Parker, ``{PRISM} 4.0: Verification of
  probabilistic real-time systems,'' in \emph{23rd International Conference on
  Computer Aided Verification (CAV)}, 2011, pp. 585--591.

\bibitem{SenVA2006}
K.~Sen, M.~Viswanathan, and G.~Agha, ``Model-checking {M}arkov chains in the
  presence of uncertainties,'' in \emph{12th International Conference on Tools
  and Algorithms for the Construction and Analysis of Systems (TACAS)}, 2006,
  pp. 394--410.

\bibitem{ter2018framework}
M.~Ter~Beek, A.~Legay, A.~L. Lafuente, and A.~Vandin, ``A framework for
  quantitative modeling and analysis of highly (re)configurable systems,''
  \emph{IEEE Transactions on Software Engineering (Early Access)}, 2018.

\bibitem{zeng2004qos}
L.~Zeng, B.~Benatallah, A.~H. Ngu, M.~Dumas, J.~Kalagnanam, and H.~Chang,
  ``{QoS}-aware middleware for web services composition,'' \emph{IEEE
  Transactions on Software Engineering}, vol.~30, no.~5, pp. 311--327, 2004.

\end{thebibliography}
\end{document}